\documentclass[useAMS,usenatbib]{mn2e}
\usepackage[letterpaper, totalwidth=480pt, totalheight=680pt]{geometry}
\usepackage{longtable}
\usepackage{graphicx}

\title[The complex case of V445~Lyr]{The complex case of V445~Lyr observed with \textit{Kepler}: Two Blazhko modulations, a non-radial mode, possible triple mode RR Lyrae pulsation, and more}
\author[E. Guggenberger, K. Kolenberg,  J. M. Nemec, R. Smolec et al.]{E. Guggenberger$^{1}$\thanks{E-mail:
elisabeth.guggenberger@univie.ac.at}, K. Kolenberg$^{2,3}$, J. M. Nemec$^{4}$, R. Smolec$^{1,5}$, J.M. Benk\H{o}$^{6}$, 
\and C.-C. Ngeow$^{7}$, J. G. Cohen$^{12}$, B. Sesar$^{12}$, R. Szab\'o$^{6}$,  M. Catelan$^{8}$, P. Moskalik$^{5}$,
\and K. Kinemuchi$^{9}$, S. E. Seader$^{10}$,  J. C. Smith$^{10}$, P. Tenenbaum$^{10}$, H. Kjeldsen$^{11}$ \\
$^{1}$Institut f\"ur Astronomie, Universit\"at Wien, T\"urkenschanzstrasse 17, A-1180 Vienna, Austria\\
$^{2}$Harvard-Smithsonian Center for Astrophysics, 60, Garden street, Cambridge MA 02138, USA\\ 
$^{3}$Instituut voor Sterrenkunde, Celestijnenlaan 200D, B-3001 Leuven, Belgium\\
$^{4}$Department of Physics \& Astronomy, Camosun College, Victoria, British Columbia, V8P 5J2, Canada\\
$^{5}$Copernicus Astronomical Center, Polish Academy of Sciences, ul. Bartycka 18, 00-716 Warszawa, Poland\\
$^{6}$Konkoly Observatory, Research Center for Astronomy and Earth Sciences, PO Box 67, H-1525 Budapest, Hungary\\
$^{7}$Graduate Institute of Astronomy, National Central University, Jhongli City, Taoyuan County 32001, Taiwan\\
$^{8}$Pontificia Universidad Cat\'olica de Chile, Facultad de F\'{i}sica, Departamento de Astronom\'\i a y Astrof\'\i sica, \\~Av. Vicu\~{n}a Mackenna 4860, 782-0436 Macul, Santiago, Chile\\
$^{9}$NASA-Ames Research Center/Bay Area Environmental Research Institute, Mail Stop 244-30, Moffett Field, CA 94035, USA\\
$^{10}$SETI Institute/NASA Ames Research Center, Moffett Field, CA 94035, USA\\
$^{11}$Department of Physics and Astronomy, Aarhus University, DK-8000 Aarhus C, Denmark\\
$^{12}$California Institute of Technology, Mail Stop 249-17, 1200 East California Boulevard, Pasadena, CA 91125, USA}

\begin{document}

\date{Accepted 0000 December 15. Received 0000 December 14; in original form 0000 October 11}

\pagerange{\pageref{firstpage}--\pageref{lastpage}} \pubyear{2002}

\maketitle

\label{firstpage}

\begin{abstract}
Rapid and strong changes in the Blazhko modulation of RR Lyrae stars, as they have recently been detected in high precision satellite data, have become a crucial topic in finding an explanation of the long-standing mystery of the Blazhko effect. We present here an analysis of the most extreme case detected so far, the RRab star V445~Lyr (KIC 6186029) which was observed with the \textit{Kepler} space mission. V445~Lyr shows very strong cycle-to-cycle changes in its Blazhko modulation, which are caused both by a secondary long-term modulation period as well as irregular variations. In addition to the complex Blazhko modulation, V445~Lyr also shows a rich spectrum of additional peaks in the frequency range between the fundamental pulsation and the first harmonic. Among those peaks, the second radial overtone could be identified, which, combined with a metallicity estimate of [Fe/H]=-2.0 dex from spectroscopy, allowed to constrain the mass (0.55-0.65 $M_{\odot}$) and luminosity (40-50 $L_{\odot}$) of V445~Lyr through theoretical Petersen diagrams. A non-radial mode as well as possibly the first overtone are also excited. Furthermore, V445~Lyr shows signs of the period doubling phenomenon and a long term period change. A detailed Fourier analysis along with a study of the O-C variation of V445~Lyr is presented, and the origin of the additional peaks and possible causes of the changes in the Blazhko modulation are discussed. The results are then put into context with those of the only other star with a variable Blazhko effect for which a long enough set of high precision continuous satellite data has been published so far, the CoRoT star 105288363.
\end{abstract}


\section{Introduction}

RR Lyrae stars, which are low-mass helium burning stars on the horizontal branch, were long thought to be rather simple radial pulsators. They follow a period-luminosity-color relation which makes them valuable distance indicators, and because of their age and evolutionary status, they are also used to study the formation and evolution of the Galaxy \citep{catelan09}. They can oscillate in the fundamental radial mode (type RRab), the first overtone (type RRc) or both modes simultaneously (type RRd), and their high amplitudes of up to 1.5 mag in V for RRab type stars made their variability easy to discover, so that they have been known since the end of the XIX century.

Already more than a hundred years ago, however, it turned out that there is an aspect to RR~Lyrae stars which is not understood at all: \Citet{Blazhko07} found a ``periodic change in the period'' of RW Dra, which he could not explain, and which still remains unexplained today. \citet{shapley16} later found in his observations of RR~Lyrae that  also the brightness of the maxima and the light curve shape show periodic changes. With increasing data quality in the recent past, the unsolved problem got even more severe, as it turned out that not just a rather small fraction of exceptional RR~Lyrae stars were affected, but probably around 40-50 per cent of all RRab stars \citep{kol10a, Benko10, jur09c}. Also among RRc type stars, amplitude and phase modulation was found to be surprisingly widespread \citep{Arellano}.

This so-called Blazhko effect was long thought to be a periodic/regular phenomenon. Traditionally, only one Blazhko period was assigned to each modulated star, and the phenomenon was expected to repeat in every Blazhko cycle, agreeing with the widely used definition that ``the Blazhko effect is a periodic amplitude and/or phase modulation with a period of several 10 to 100 pulsation periods". There were several reports about changes in the Blazhko modulation of various stars (see section 5 of \citet{gug11} for a recent summary), but those reports usually had to rely on sparse data with large gaps, so that it was impossible to say when exactly a change took place and whether it happened continuously or abruptly. The Blazhko effect was therefore still considered to be a strictly repetitive phenomenon with only some rare exceptions showing secondary modulation periods \citep[for example CZ Lac,][]{sodor11} or changes on very long time scales. It was not until the availability of ultra-precise data from space missions like CoRoT that strong and irregular cycle-to-cycle changes of a Blazhko star were documented and that it became obvious that seemingly chaotic phenomena need to be accounted for when modelling the Blazhko effect.

While the detection of cycle-to-cycle changes in the Blazhko modulation posed a significant challenge for all classical models that required a clock-work-like behaviour, some new ideas were published. \citet{sto06} suggested that  transient small-scale magnetic fields modulate the turbulent convection inside the helium and hydrogen ionisation zones, a mechanism which certainly could explain subsequent Blazhko cycles of differing strength. This scenario, however, was recently tested on the basis of hydrodynamical models by \citet{smo11} who found that it was not possible to reproduce all observed properties of the light curve, even when allowing a huge modulation of the mixing length. On the other hand, \citet{buchler11} successfully modeled both regular and irregular modulations by using the amplitude equation formalism. In their models, a strange attractor in the dynamics causes chaotic behaviour. The 9:2 resonance between the fundamental mode and the 9th overtones that was found by \citet{kollath11} to be the reason for the recently discovered phenomenon of period doubling \citep{szabo10}, seems to play an important role in causing a Blazhko modulation.

We present here a study of the most extreme case detected so far: V445~Lyr (KIC~6186029). This RRab star not only shows the strongest cycle-to-cycle changes found up to now, but also a rich spectrum of additional modes in the region between 2 and 4 $\rm{d}^{-1}$. Both Fourier and O-C analyses are used to investigate the variability of the pulsation and of the modulation (Section \ref{fourier}), and the results are compared to the case of CoRoT~105288363 in Section~\ref{comp-corot}. Spectroscopy as well as theoretical Petersen diagrams are used to determine the fundamental parameters like metallicity, luminosity and mass (Section \ref{param}). Additionally, the new \textit{analytic modulation approach} for data analysis recently proposed by \citet{benko11} is applied to the data in Section \ref{analytical}.

\section{Background information on V445~Lyr}
\label{target}
V445~Lyr, with the coordinates RA 18h 58m 26s and Dec 41$^\circ$ 35' 49"  (J2000) is also known as KIC~6186029, or GR244, and has a \textit{Kepler} magnitude of \textit{Kp}=17.4. Two publications from the pre-\textit{Kepler} era exist for this target: \citet{romano72} found it to be variable and classified it as an RR~Lyrae, and \citet{kukarkin73} included it into his name list of variable stars. \citet{romano72} also lists the photographic brightness of maximum and minimum to be 15.3 and 17.3 mag, respectively, indicating a surprisingly large amplitude of 2 mag, much higher than even during extreme Blazhko maxima in the modern data. This might at least partly be explained, however, by the difference between the observed bandpasses. Unfortunately, no details of the observations and no light curves are given, and the forthcoming paper that was announced by the author could not be found. We therefore cannot know if the observed amplitude of 2 mag is real or due to possible observational errors. No error estimations of the observations were given by \citet{romano72}.

Since \textit{Kepler} data have become available, two more publications have dealt with V445~Lyr, both presenting the \textit{Kepler} data up to Q2: \citet{szabo10} listed V445~Lyr as a possible candidate for the period doubling phenomenon, and \citet{Benko10} already noted changes in the Blazhko modulation of V445~Lyr, and reported the presence of radial overtones.

\section{\textit{Kepler} photometry}

\begin{figure*}
\includegraphics[width=170mm, bb= 0 0 594 385]{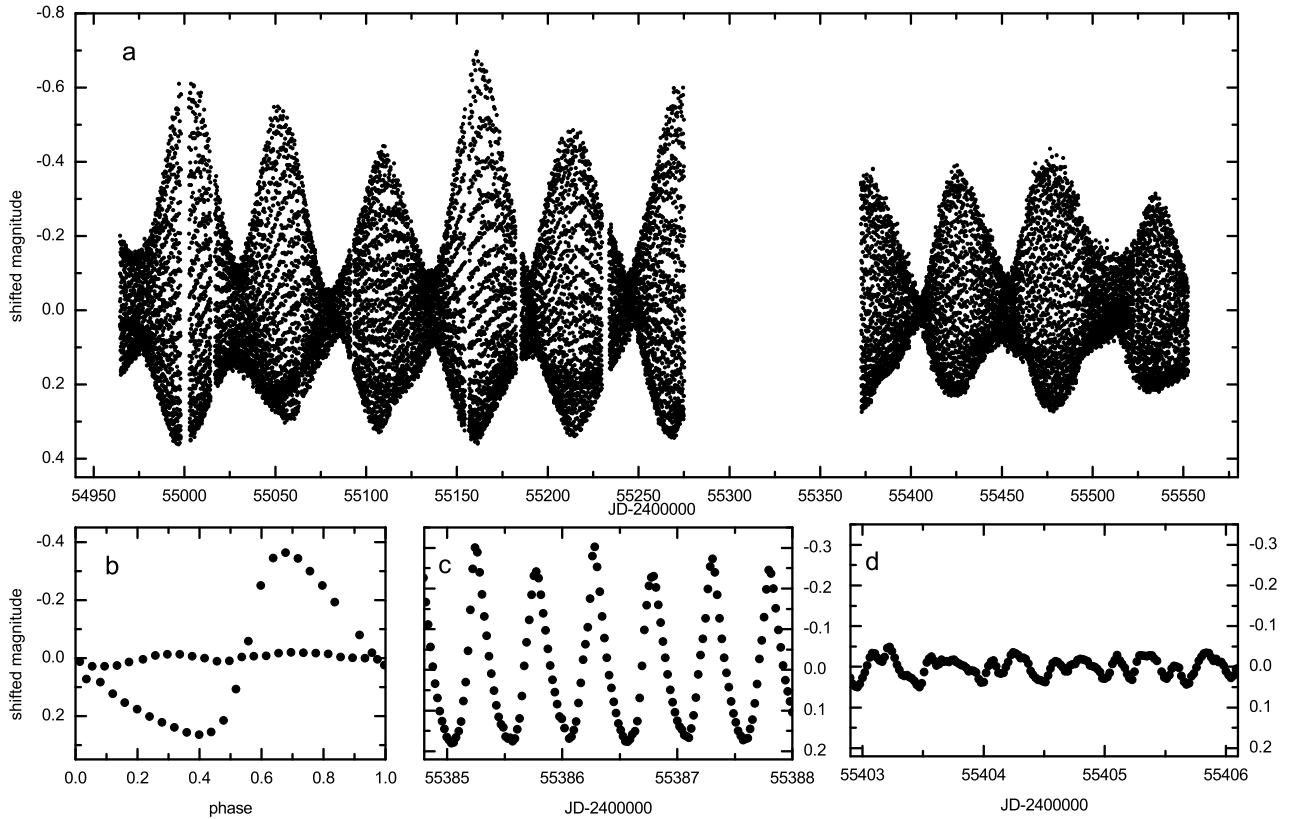}
  \caption{Panel a shows the complete light curve of V445 Lyr presented in this paper, including quarters Q1-Q7. Due to failure of a module, there are no data from Q5, resulting in the gap from MJD 55273 to 55372. Panel b illustrates the extreme Blazhko modulation by comparing a pulsation cycle at Blazhko maximum to one at Blazhko minimum. The light variation has an extremely low amplitude during this Blazhko minimum and shows a double maximum. In panel c, a zoom into a region with period doubling is shown as an example, and panel d emphazises the distorted light curve shape that occurs during some Blazhko minima.}
\label{fig-lightcurve}
\end{figure*}

The \textit{Kepler} space mission was launched on March 6, 2009 into an Earth-trailing heliocentric orbit \citep{koch10}. Its primary purpose is the detection of Earth-sized planets in the habitable zone of solar-like stars through the transit method, which requires continuous and ultra-precise photometry of over 150000 stars for at least 3.5 years. This is also the duration of the primary mission. \textit{Kepler} therefore provides not only the longest continuous data sets ever observed for RR Lyrae stars, but also does so with the highest photometric precision ever obtained, as a consequence greatly improving our knowledge about stellar pulsations.

The \textit{Kepler} spacecraft carries a Schmidt telescope with an aperture of 0.95~m and 42 science CCDs which cover a field of view of about 115 square degrees \citep{jen10}. The photometric bandpass ranges from 423 to 897~nm, thus avoiding the CaII H and K lines in the blue, and fringing due to internal reflection in the red. The \textit{Kepler} band is therefore slightly broader than a combination of Johnson V and R, and \textit{Kepler} magnitudes are usually about 0.1 mag from R for most stars \citep{koch10}.

Every quarter orbit, the spacecraft is rotated in order to keep the solar panels oriented towards the Sun, and the radiator which cools the focal plane towards deep space \citep{haas10}. Data from different quarters are denominated Q1, Q2, etc. Each of the quarters which are used here has a time base of about 90~d, except Q1 which covers about 33~d. 

Each measurement is based on a 6.02~s exposure plus a 0.52~s readout time. To obtain Long Cadence data (LC, 29.4~min), 270 measurements are coadded; for Short Cadence data (SC, about 1~min), 9 exposures are coadded \citep{christ}. 

The time of each measurement is given in truncated barycentric Julian Date (HJD-2400000), and refers to the midpoint of the measurement.\\

\subsection{The V445 Lyr data set}
Figure \ref{fig-lightcurve} illustrates the data obtained for V445~Lyr. In this paper, we present LC data obtained in the quarters 1 to 7, with a total time base of 588~d. Because of the loss of module 3, which happenend in January 2010, there are no data available for this target in Q5. Some smaller gaps are also present in the data as can be seen in Fig.~\ref{fig-lightcurve}. Some of them are due to unplanned safe mode events or loss of fine point control, others are caused by the regular downlinks where the spacecraft's antenna is pointed towards Earth for data transmission, and science data collection is interrupted. 

\textit{Kepler} provides light curves in two different formats: raw flux or corrected flux. The latter has been processed for planet transit search by the PDC (Pre-search Data Conditioning) pipeline which is known to sometimes remove astrophysical features and which does not preserve all stellar variability. Hence here, and for the study of variable stars in general, we make use only of the raw time series.

As the spacecraft is rotated every quarter orbit, the target falls on different CCDs after each ``roll'', resulting in differences in average flux due to different sensitivity levels. Additionally, trends might occur due to image motion on the CCD or sensitivity changes. The scaling and detrending which has to be performed before starting the analysis is a delicate task especially for targets like RR Lyrae stars which have high amplitudes, long periods and changes in the amplitudes (see \citet{celik11} for a detailed discussion). Here, we removed  linear trends which were determined from a running average separately for every quarter, and scaled every quarter to the same mean brightness. The continuity of the upper and lower envelope of the lightcurve was a good indicator of correct scaling. Note that due to the gap during Q5, the continuity could not be checked there. For this reason, the scaling of the data obtained after the gap might not exactly be consistent with the scaling applied before Q5.

The scaled and detrended data that were used in this analysis are available as online material in the format displayed in Table~\ref{datatable}.
\begin{table}
\caption{The scaled and detrended data set of V445~Lyr that was used for the analysis in this paper. Column 1 gives the truncated barycentric Julian date, column 2 the magnitude with the average shifted to zero, and column 3 gives the quarter in which data were obtained. The full table is available online only.}
\begin{center}
\begin{tabular}{lcc}
\label{datatable}
HJD-2400000 & mag & quarter \\
\hline
54964.51211   &     0.044188798   &     Q1 \\
54964.53254   &     0.038912686    &    Q1\\
54964.55298   &     0.017979982    &    Q1\\
54964.57341   &     -0.023960246   &     Q1\\
54964.59385   &     -0.099222117    &    Q1\\
... & ...& ...\\
\hline
\end{tabular}
\end{center}
\end{table}

Panel a of Fig. \ref{fig-lightcurve} shows the complete combined data set of V445~Lyr. A total of 5.5 and 3.5 Blazhko cycles were observed before and after the large gap, respectively, and 900 pulsation cycles were observed. Already from looking at panel a it is obvious that the Blazhko effect shows strong and fast changes, with almost all observed cycles having a different appearance. Panel b compares in a phase diagram directly a light curve from a Blazhko maximum with one obtained during the subsequent Blazhko minimum, illustrating the extremely low pulsation amplitude that occurs during some, but not all, modulation minima. Note that due to the very small amplitude and the distorted light curve showing a double maximum (see also panel d of Fig.~\ref{fig-lightcurve}) it would not be possible to recognize the RR Lyrae nature of this star when observing it only during Blazhko minimum where the peak-to-peak amplitude can be as low as 70~mmag. Extremely low pulsation amplitudes during Blazhko minima were recently also reported by \citet{Sodor12} for two of RRab type stars, and the occurrence of a strong bump during Blazhko minimum was noted in RZ~Lyr \citep{jur12}. In panel c of Fig.\ref{fig-lightcurve}, a small part of the light curve around JD 2455386 is shown, illustrating the phenomenon of period doubling (alternating higher and lower light maxima) which was recently discovered in \textit{Kepler} Blazhko stars \citep{szabo10} and which is also present in V445 Lyr.

\section{Analysis and Results}
\label{fourier}

\begin{figure}
\includegraphics[width=90mm, bb= 0 0 440 360]{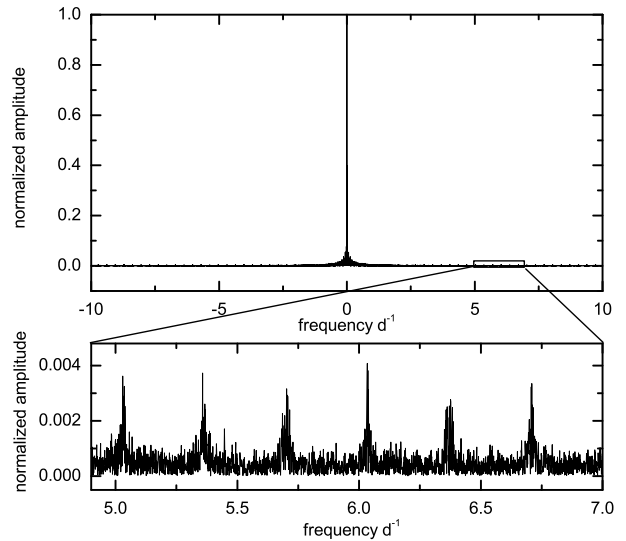}
\caption{Spectral window function of the complete V445 Lyr data set. Lower panel: magnification of the region between 5 and 7 $\rm{d}^{-1}$, showing some examples of the low-amplitude features caused by the reaction wheel.}
\label{fig-sw}
\end{figure}

Due to the quasi continuous coverage and the high photometric precision, the conditions for a Fourier analysis are more than favorable. The spectral window function (see Fig.~\ref{fig-sw}) is almost perfect, without any alias peaks visible at first glance. Because of the Earth-trailing orbit, no orbital frequencies like the ones that cause trouble in the data from many other space missions are present here.

A well-known feature in the \textit{Kepler} data are the momentum desaturations of the reaction wheel, which happen every 2.98~d. During the thruster firings which are necessary to release the angular momentum that has built up in the reaction wheels, the space craft momentarily loses fine point control \citep{vancleve}. This results in a missing data point during every desaturation in the otherwise regularly spaced data, leading to a comb-like structure in the Fourier transform with a spacing of 0.335~$\rm{d}^{-1}$. In the spectral window, those peaks have a very low normalized amplitude of only 0.004 mag. A zoom into the spectral window (see lower panel of Fig.~\ref{fig-sw}) reveals this comb of tiny peaks. When carefully removing all high-amplitude peaks from the Fourier spectrum before turning to interpreting features with amplitudes which are orders of magnitude smaller as are discussed in section~\ref{addmodes}, those alias peaks are not expected to cause any trouble.

Other features in the spectral window function are two harmonics of the monthly data downlink frequency, at 0.065 and 0.13 $\rm{d}^{-1}$, with normalized amplitudes of 0.05 and 0.03 mag, respectively.

Due to the time base of 588~d, the Rayleigh frequency resolution is 0.0017 $\rm{d}^{-1}$.

\subsection{Fourier analysis of the full data set}
\label{fulldata}
As a first step, a Fourier analysis of the complete data set was performed, keeping in mind that changes of the fundamental period might occur during the time of the observations, and that the changes of the Blazhko modulation cause a large number of peaks close to the classical pulsation and modulation components. The Fourier analysis of the full data set, however, is necessary to obtain an overall picture of the pulsation and modulation properties of V445~Lyr,  to find the mean values of the pulsation and modulation periods and to detect possible long-term periodicities which can be resolved with a long time base only.

The Fourier analysis of the complete data set was performed with Period04 \citep{Lenz} and then checked with SigSpec \citep{reegen07}. The results agreed within the errors. The Fourier analysis revealed a mean pulsation period of $0.513075 \pm 0.000005 $ d and a mean Blazhko period of $53.1 \pm 1$~d with the ephemeris 
\[\rm{HJD}~(T_{\rm{max, pulsation}})= 2455550.514 + 0.513075 \times \rm{E_{pulsation,}}\]
\[\rm{HJD}~(T_{\rm{max, Blazhko}})= 2455534.2 + 53.1 \times \rm{E_{Blazhko.}}\]
Note that the presence of close peaks has a great influence on the fitting and frequency optimization procedure, and the errors are therefore larger than would normally be expected for a Fourier fit to data of the given quality.

\subsubsection{Multiplet components}
The Blazhko multiplets (i.e., the pattern of peaks which is typical for Blazhko stars with peaks at the positions $kf_{0} \pm nf_{B}$, with $k$ and $n$ being integers denoting the harmonic order and the multiplet order, respectively, and with $f_0$ and $f_B$ denoting the fundamental and the Blazhko frequency) were found to be very asymmetric in amplitude. Much higher amplitudes appeared on the right (higher frequency) side than on the left. This is the more common case, which is observed in three fourths of all Blazhko RRab stars \citep{alcock03}. Components were detected up to quintuplet order on the right side, while on the left side of the main pulsation component, only one side peak (i.e., a triplet component, $n=1$) could be found. In some orders, peaks were found near the positions that would be expected for septuplets, but the deviations from the exact frequency values were considered too big to identify the peaks safely as septuplet components.

In addition to the classical Blazhko multiplets around the fundamental mode and its harmonics, also a secondary modulation component, hereafter $f_S$, could be identified. It manifests itself as a series of additional peaks next to the classical modulation sidepeaks of the Fourier spectrum, appearing in every order with a spacing of $0.00698 \pm 0.00027~\rm{d}^{-1}$, indicating a secondary modulation period of $143.3 \pm 5.8$ d. The secondary peaks appearing on the right side of the classical peaks show surprisingly high amplitudes, while the secondary peaks to the left of the classical multiplet have small amplitudes and could only be detected after prewhitening a large number of higher amplitude peaks. Interestingly, the peaks belonging to this additional series could also be detected in higher multiplet orders ($n=3$) than the classical multiplet, making it difficult to explain the peaks as combination frequencies. Their amplitudes seem to decrease less quickly with increasing harmonic order $k$ (see also section~\ref{decr}), making them easier to spot at high orders.

\begin{figure}
\includegraphics[width=90mm, bb= 0 0 285 600]{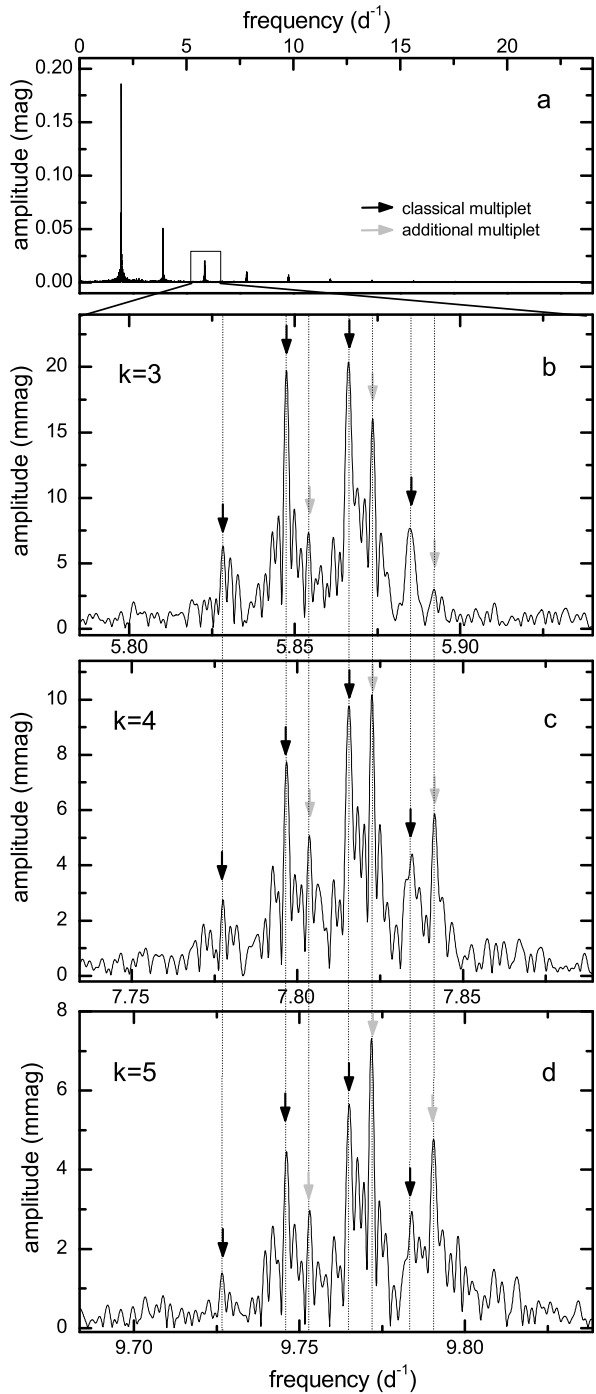}
\caption{Fourier transform of the complete V445~Lyr data set (upper panel). In the lower panels (b, c and d), the regions around the harmonics $kf_0$ with k=3, 4 and 5 are shown in detail. Black arrows indicate the peaks belonging to the classical Blazhko multiplet with $kf_0-f_B$, $kf_0$, $kf_0+f_B$ and $kf_0+2f_B$, while grey arrows indicate some of the peaks belonging to the secondary multiplet which within itself has a spacing equal to the Blazhko frequency, but is shifted with respect to the classical multiplet by $0.00698~\rm{d}^{-1}$. Note that from order k=4 onwards, the highest amplitude peak in this order is a peak belonging to the secondary multiplet}.
\label{fig-fourier}
\end{figure}

Figure~\ref{fig-fourier} shows the Fourier transform of the data, providing also zooms into the regions around the $3^{rd}$, $4^{th}$ and $5^{th}$ harmonic order where both the classical multiplet and the additional components can be seen. The highest peaks among the multiplet components, which can be seen even before prewhitening the original data, are marked with arrows. 

\begin{figure}
\includegraphics[width=90mm, bb= 0 0 450 450]{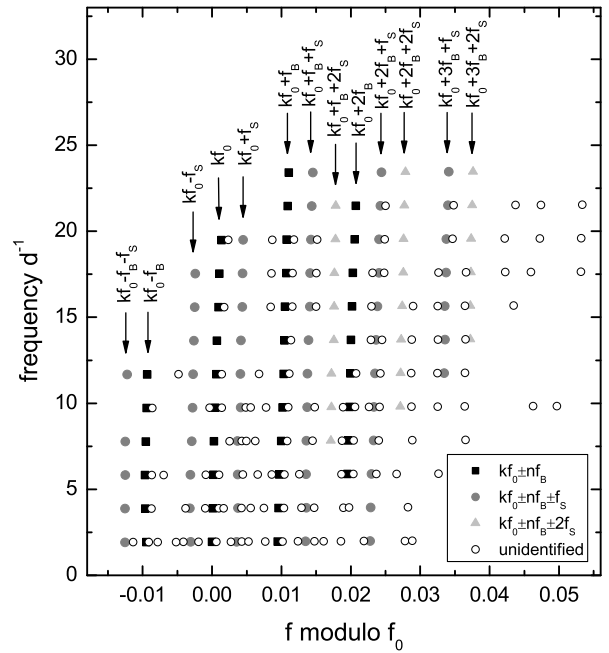}
\caption{``Echelle'' diagram of the peaks detected in the vicinity of the fundamental pulsation mode and its harmonics. Frequency is plotted versus $f$ modulo $f_0$, so that regularly spaced peaks which appear in every order align in vertical patterns, making it easy to identify combinations with $f_0$. Peaks belonging to the category $kf_0 \pm nf_B$ are shown as black squares, peaks of the group $kf_0 \pm nf_B + f_S$ are shown as grey circles, components of $kf_0 +nf_B +2f_S$ are plotted with light grey triangles. Open circles denote unidentified and/or unresolved peaks, some of which originate from the non-repetitive nature of the modulation and long-period changes.}
\label{fig-echelle}
\end{figure}

Figure~\ref{fig-echelle} illustrates the pattern of detected frequencies in the vicinity of the fundamental pulsation and its harmonics in the style of an ``echelle'' diagram. In this diagram, which is similar to the diagrams used to unveil equally spaced peaks in helioseismology, the frequency of each peak is plotted against ($f$ modulo $f_0$), i.e., $f/f_0-INT(f/f_0)$, or $f/f_0-INT(f/f_0)-1$ for peaks to the left of the harmonic, therefore clearly revealing patterns which repeat in every harmonic order. Peaks belonging to the same group of combinations align in vertical ridges. We stress that unlike in the helioseismic application, where the ridges denote different radial orders of same degree, in this case the echelle diagram only serves the purpose of displaying in a very practical and easy way the repeating patterns in different harmonic orders of non-sinusoidal fundamental radial pulsation.

\subsubsection{Deviation of the harmonics}
\label{dev}
Due to the non-sinusoidal light curve shape typical for RR Lyrae stars, harmonics of the fundamental mode are expected to appear at the frequencies $kf_0$, where $k$ is an integer denoting the harmonic order. The classical Blazhko multiplets in modulated stars are spaced equidistantly, implying frequency values of $kf_0 \pm nf_B$ with $n$ denoting the multiplet order. Long-term period changes and close peaks caused by irregular phenomena, however, can distort this frequency pattern. When analyzing time series of Blazhko RR~Lyrae stars, there are two options for fitting the data: one is to fix the frequencies of the harmonics and Blazhko multiplets to their expected values of $kf_0 + nf_B$, therefore reducing the number of free parameters in the fit. The other option is to let all parameters, including the frequency values, free. When the latter option was applied to this data set, the harmonics were observed to deviate systematically and significantly from their expected values, which can also be noticed as a slight rightwards tilt of the ridges in the echelle diagrams (Fig.~\ref{fig-echelle}). Normally, one would expect that this is simply caused by a wrong value for $f_0$, but in this case, no value of $f_0$ could be found which could solve the issue, i.e., every detected harmonic, when divided by its order, required a different $f_0$. We therefore decided in favour of the more pure approach and did not fix the frequency values to the expected positions, but left all parameters free in the fit of the complete data set. The problem disappeared, however, when analyzing different subsets of data separately (see section~\ref{subsets}), and we therefore suspect it to be either the result of period changes which take place during certain seasons (see also section~\ref{OC}) and/or close unresolved peaks which are known to strongly influence the results of both the Fourier analysis and the multisine fitting procedure.\\

\subsubsection{Amplitudes versus harmonic order}
\label{decr}
\begin{figure}
\includegraphics[width=90mm, bb= 0 0 450 440]{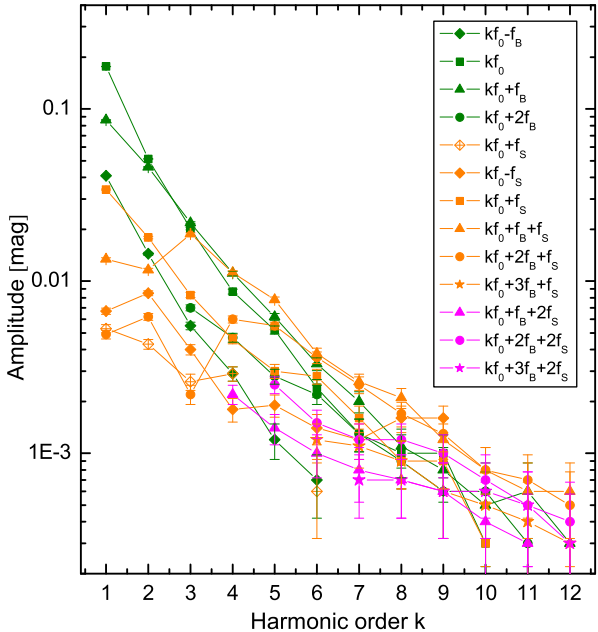}
\caption{Amplitudes of all components versus harmonic order. Components belonging to the classical Blazhko multiplet are shown with green symbols, peaks which are combinations with $f_S$ are shown in orange, and components with a $2f_S$ term are shown in magenta. It can be seen that the amplitude of the $f_S$ multiplets decrease less steeply than those of the classical Blazhko multiplet, reaching the same amplitude in the 4th order and dominating the frequency spectrum for higher orders.}
\label{fig-ampdecr}
\end{figure}

It is a well-known fact that in Blazhko RRab stars the amplitudes of the multiplet side peaks decrease less rapidly with harmonic order than that of the main component. This was first described by \citet{jur05}, and then confirmed for other well-studied stars like SS~For \citep{kol09}, RR~Lyr \citep{kol10b} and CoRoT~105288363 \citep{gug11}. As expected, the result is the same for V445~Lyr (see Fig.~\ref{fig-ampdecr}), but in an extreme way with the ampitude of the right side peak exceeding that of the main component as early as in the 3rd order. Additionally, the amplitudes of the secondary modulation multiplet, i.e. combinations with the secondary modulation $f_S$, could be studied in this case. It turned out that the amplitudes of the secondary multiplet components decrease even less steeply, therefore dominating the Fourier spectrum from the 4th order onwards. The strongest signal then comes from the combination $f_0+f_B+f_S$, i.e. the peak on the right side of the right triplet component. Also, combinations with the term $2f_S$ could be detected, and their amplitudes are also plotted in Fig.~\ref{fig-ampdecr}. They also show a very slow amplitude decrease.\\

\subsubsection{Number of relevant frequencies}
\label{number}
Due to the dense spectrum of peaks which is caused by the cycle-to-cycle changes of the Blazhko effect, an analysis in the classical sense, i.e. taking into account all frequencies down to a certain S/N level or a certain significance criterion, might not be optimal in a case like V445 Lyrae, as it does not yield meaningful results. A large number of the detected peaks is likely to be the result of ``stellar noise'', caused by irregular and/or long-periodic phenomena, and many of them are not resolved with the available time span. Tests revealed that as many as 771 frequencies can be found when performing an analysis until the generally adopted criteria of significance are reached. Many of them were not resolved, and many could not be attributed to any combination of other modes, and did not show repeating patterns in the echelle diagrams. Therefore, instead of choosing the classical approach, the analysis was stopped after a certain number of the highest peaks in every harmonic order was found and prewhitened, usually around 20 peaks per order. It turned out that after subtracting approximately 20 peaks in a given harmonic order, no meaningful combinations could be identified among the following peaks, and many unresolved peaks appeared. In Fig.~\ref{fig-echelle} only the highest peaks of every order are shown, already including some unresolved peaks which could not be avoided due to their high amplitude. As the Nyquist frequency of LC data is 24.4 $\rm{d}^{-1}$, 12 harmonic orders could be observed, and 239 frequencies were subtracted around the main pulsation components until the attention was turned towards the additional peaks which are present in the region between the harmonics of the fundamental mode (see next section). 

Altogether, 239 frequencies were included, of which one is the fundamental mode, 9 are harmonics of the fundamental mode, 104 are combinations of $f_0$ or its harmonics with the Blazhko frequency $f_B$ and/or the secondary modulation $f_S$, and 125 are unidentified. \\

\subsubsection{Additional frequencies: radial overtone and non-radial pulsation}
\label{addmodes}
It was already noted by \citet{Benko10} that V445~Lyr shows a rich spectrum of frequencies in the region between the harmonics of the fundamental mode. Those frequencies are not at all typical for ab-type RR~Lyrae stars and have never been detected in such a large number in an RR Lyrae star. Some peaks were suspected to be radial overtones by \citet{Benko10}, and also frequencies which are half-integer multiples of the fundamental mode are expected to appear in this region as a consequence of the period-doubling phenomenon as described by \citet{szabo10}. But there is more than this to V445~Lyr.

\begin{figure}
\includegraphics[width=90mm, bb= 0 0 535 380]{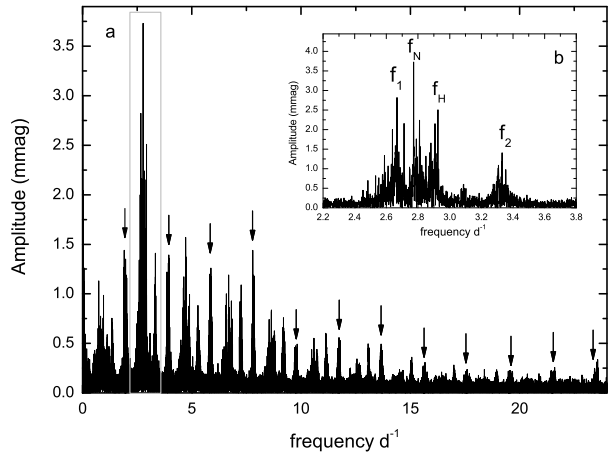}
\caption{Fourier spectrum after prewhitening 239 frequencies in the vicinity of the fundamental mode and its harmonics. Very clearly, additional frequencies can be seen between the harmonic orders, as well as their combinations with $f_0$ which are present up to high harmonic orders. Arrows mark the places where the harmonics and the Blazhko as well as the secondary multiplets were located before prewhitening. Some signal remains around their positions, as discussed in section~\ref{number}. The insert (panel b) shows a zoom into the region between the fundamental mode and the first harmonic (2-4 $\rm{d}^{-1}$), indicated as a grey square.}
\label{fig-addfourier}
\end{figure}

Fig.~\ref{fig-addfourier} shows the frequency spectrum after subtraction of the relevant peaks around the multiples of the fundamental pulsation as discussed in the previous section (their former positions are marked with arrows).  The additional frequencies can clearly be seen to be the dominant signal with an amplitude of 3.7 mmag for the highest peak. The region between the fundamental mode and its first harmonic is indicated with a grey box and enlarged in the insert (panel~b). Four frequency regions with enhanced signal can be noted in the enlargement: around 2.65, 2.8, 2.9 and 3.33~$\rm{d}^{-1}$. This pattern repeats in every harmonic order, indicating combinations of the frequencies with the fundamental mode and its surrounding peaks. The presence of combinations is a strong evidence that the additional signal is not introduced by a possible background star.

\begin{figure}
\includegraphics[width=90mm, bb= 0 0 525 380]{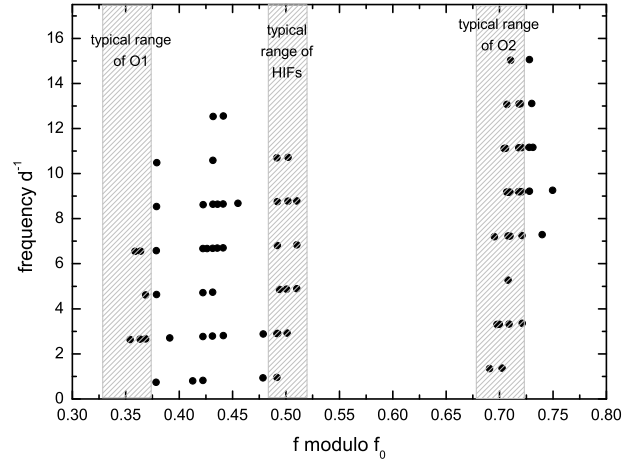}
\caption{Echelle diagram of the additional frequencies, showing how the patterns repeat in every harmonic order. Shaded boxes indicate the ranges which would be typical for the first and second overtone (O1 and O2, respectively), as well as for the half-integer frequencies (HIFs).}
\label{fig-addechelle}
\end{figure}

In a Fourier analysis of the relevant frequency regions, 80 peaks (including combinations) were considered significant and subtracted. A closer inspection of the result revealed that the dominant peaks formed combinations not only with the fundamental mode but also with the Blazhko multiplet peaks (including quintuplets!) and in some cases also with the peaks belonging to the secondary multiplet. Also negative combinations such as $f_N-f_0-f_B$ occur. Significant combination peaks can be traced up to the fifth harmonic order, but an excess in signal is visible in the Fourier spectrum even at much higher orders (see Fig.~\ref{fig-addfourier}). In Fig.~\ref{fig-addechelle}, an echelle diagram is plotted for the additional peaks, clearly showing the combinations with the fundamental mode aligned in vertical patterns. Shaded boxes indicate the typical regions in which overtone modes and half-integer combination frequencies (HIFs) would be expected to be situated. Please note that it was shown by \citet{szabo10} that due to the on- and offset of the period doubling phenomenon, the HIF peaks are not necessarily located at the exact positions of the half-integer multiples, but might deviate by several per cent. Therefore, the shaded box at ($f$ modulo $f_0$)=0.5 in Figure~\ref{fig-addechelle} has a distinct width.

In V445~Lyr, the frequency at 2.9256 $\rm{d}^{-1}$ deviates by only 0.0021 $\rm{d}^{-1}$ (i.e., 0.07 per cent) from the exact value of $3f_0/2$. Given the fact that clear signs of period doubling are indeed visible in the light curve (see panel c of Fig.~\ref{fig-lightcurve}), and considering the above-mentioned findings of \citet{szabo10}, it is quite safe to interpret this frequency as an HIF caused by period doubling. We hereafter refer to it as $f_H$. Four significant HIFs (which can also be interpreted as combinations of $f_H$ with $f_0$) were found in the Fourier spectrum: $3f_0/2$,  $5f_0/2$,  $9f_0/2$, and $11f_0/2$. Combinations with the Blazhko frequency, both positive and negative, could also be identified (see Table~\ref{tab-addmodes} for a complete list).\\

Another interesting feature is the peak at 3.3307 $\rm{d}^{-1}$ which shows a frequency ratio of $f_0/f_2$=0.585 with the fundamental mode, and which we hereafter refer to as $f_2$. Its period ratio is typical for the second overtone. Peaks with similar period ratios have already been reported for several RR Lyrae stars. The first to find them were \citet{poretti} in the CoRoT star 101128793. The frequency ratio in their study was 0.584 with the fundamental mode at $f_0=2.11895~\rm{d}^{-1}$. They interpreted the peak as the second radial overtone. The same authors also reanalysed the data of V1127~Aql \citep{cha} and MW~Lyr \citep{jur08} and found frequency ratios of 0.582 ($f_0=2.8090~\rm{d}^{-1}$) and 0.588 ($f_0=2.5146~\rm{d}^{-1}$), respectively. In the sample of \textit{Kepler} RRab stars, \citet{Benko10} reported the presence of the second overtone in four different stars: V354~Lyr (a Blazhko star with $f_0=1.78037~\rm{d}^{-1}$), V2178~Cyg  (a Blazhko star with $f_0=2.05423$), V445~Lyr (the subject of this paper) and the non-modulated RRab star V350~Lyr ($f_0=1.68282$) which was the first example of a non-Blazhko double mode RR Lyrae star with the fundamental (F) and the second overtone (O2) excited. \citet{gug11} found evidence for the second overtone in CoRoT 105288363, a Blazhko star with rapid changes in the Blazhko effect, and \citet{nemec} found the second overtone in KIC~7021124, therefore providing another example of a non-modulated RRab star pulsating in both F and O2 ($f_0=1.606445~\rm{d}^{-1}$, $f_0/f_2=0.593$).

It is interesting to note how different the stars are for which the second overtone has been documented so far: Their fundamental frequencies range from about 1.6 to 2.8 $\rm{d}^{-1}$, covering almost the full bandwidth of RRab pulsation, and with respect to stability they range from non-modulated stars with almost perfectly regular RRab pulsation (V350~Lyr, KIC~7021124) to Blazhko stars with a rather regular Blazhko effect (CoRoT~101128793 and V1127~Aql) and finally to modulated stars that show dramatic changes of their Blazhko modulation (CoRoT~105288363 and V445~Lyr). Also, they cover a significant range of Blazhko periods, from 16.6 d to more than 200 d, as estimated for V2178~Cyg.

\begin{figure}
\includegraphics[width=90mm, bb= 0 0 515 320]{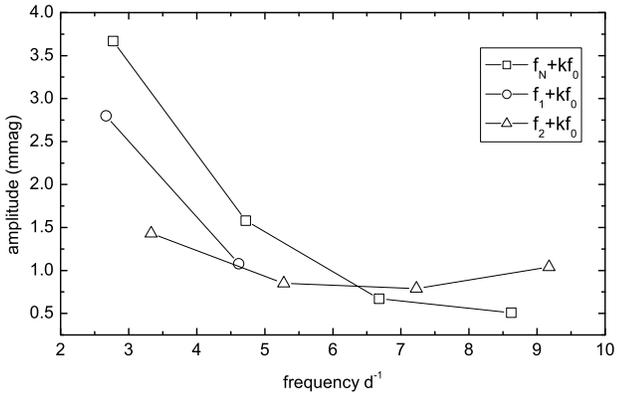}
\caption{Amplitudes of the additional frequencies and their combinations with $f_0$ versus frequency. While the amplitudes of $f_N$ and $f_1$ decrease exponentially with harmonic order, the amplitude of $f_2$ remains almost stable. Amplitude errors are smaller than the symbols.}
\label{fig-addamp}
\end{figure}

The combinations of $f_2$ in V445~Lyr deserve some special attention. While as many as 32 peaks are detected near the positions of $f_2+kf_0$, and clearly an excess of signal is visible in every harmonic order (see Figs. \ref{fig-addfourier} and \ref{fig-addechelle}), it was not possible to identify most of the detected peaks as exact combinations with the known frequencies. For $f_H$ (which was discussed in the previous paragraph), 12 out of 16 peaks could be attributed to combinations while for $f_2$, only 3 combinations with $f_0$ were found at their exact positions. All the other peaks in the vicinity of the combinations deviated too much from the calculated values to be safely matched with combinations. This is especially remarkable as the amplitudes of those peaks are surprisingly large in higher harmonic orders compared to the combinations of the other additional frequencies. From Fig.~\ref{fig-addfourier} it is obvious that in the first harmonic order, $f_2$ has a small amplitude compared to the other additional frequencies, while at the orders 4-5 they become equal, and at higher orders, the peaks in the area around $f_2+kf_0$ are the dominant feature. This is also illustrated in Fig.~\ref{fig-addamp}, where the amplitudes of the safely identified peaks are shown versus frequency. The large number of unidentified peaks around $f_2$ might indicate that the amplitude of $f_2$ is variable, either irregularly or on a timescale other than the Blazhko frequency. This will be discussed in more detail in Section~\ref{f2stability}.\\

The highest-amplitude peak among the additional frequencies (3.7 mmag) is the one at 2.7719 $\rm{d}^{-1}$. This peak was interpreted as the first overtone by \citet{Benko10}, but its ratio with the fundamental ($f_0/f_N$=0.703) is very low compared to the canonical value of 0.74-0.75. Note that the OGLE III RR Lyrae stars in the Galactic bulge have $f_0/f_1$ values going down to about 0.726 only \citep[see][]{Soszynski11}. Extremely high metallicity values would be necessary according to models \citep{smolec08, popielski, szabo04} to fit this frequency ratio with the first overtone. New spectroscopic results revealed, however, that the metallicity of V445~Lyr is [Fe/H]=-2.0 $\pm$ 0.3 (see Section~\ref{param}), rendering it impossible to explain this frequency with a radial overtone mode. $f_N$ is therefore most likely to be a non-radial mode. We note that \citet{vanhoolst} found in their non-adiabatic non-radial calculations the excitation of non-radial modes in the vicinity of the radial mode in RR Lyrae stars to be very likely, and \citet{DC99} noted in their model survey the presence of strongly trapped non-radial modes with very high growth rates near the first overtone.

\begin{figure}
\includegraphics[width=90mm, bb= 0 0 575 340]{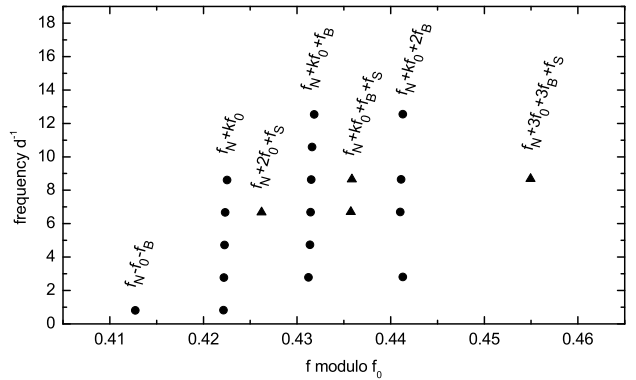}
\caption{Detail of the echelle diagram, showing only the vicinity of the mode $f_{\mathrm N}$. All detected peaks could clearly be identified as combinations of $f_{\mathrm N}$ with either the classical Blazhko multiplet or the secondary multiplet. The fact that no unidentifiable peaks are among the highest ones points towards a very stable amplitude of this mode, compared to the other additional peaks that were found in this star.}
\label{fig-addcombi}
\end{figure}

Among the additional modes in V445~Lyr, $f_N$ is the one that shows the clearest pattern of combination frequencies: of 20 peaks which were found significant in relation to $f_N$, all 20 could be unambiguously identified as exact combinations with $f_0$, the Blazhko frequency and the secondary modulation frequency (see also Table~\ref{tab-addmodes} and Fig.~\ref{fig-addcombi}, which illustrates the regular pattern of combination peaks). We note that the possible non-radial mode which was found by \citet{cha} in V1127~Aql has a very similar frequency ratio (0.696), maybe hinting at a possible systematic preference in non-radial mode excitation in RR Lyrae stars.\\

There remains, however, the fourth region of increased signal with a main peak at 2.6676 $\rm{d}^{-1}$, which, with a frequency ratio of $f_0/f_1$=0.731 is in principle in the possible range of the first overtone pulsation. One has to note that double mode RR~Lyrae stars usually follow a well-defined empirical sequence in the Petersen diagram, in other words, there is a relation between $f_0$ and $f_0/f_1$ \citep[see][]{popielski, so09}. If the peak at 2.6676 $\rm{d}^{-1}$ is indeed the first overtone, V445~Lyr would be an exception to this relation, which is very unlikely. On the other hand, outliers from the sequence similar to V445~Lyr have recently been reported by \citet{Soszynski11} in the OGLE III survey of the Galactic Bulge.

\begin{figure}
\includegraphics[width=90mm, bb= 0 0 585 400]{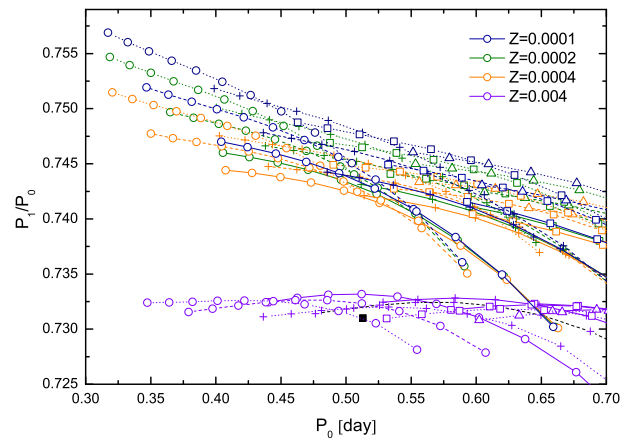}
\caption{Petersen diagram for the peak at 2.6676 $d^{-1}$ illustrating that a metallicity of Z=0.004 would be necessary to identify this peak as the first radial overtone mode. Models around the spectroscopically derived value of Z=0.0003 (see Section~\ref{param}) are also shown for comparison. Different symbols indicate different luminosities, and different line styles indicate different masses (see also Section~\ref{param}). Models were calculated with the Warsaw codes \citep{smolec08}.}
\label{Fig-O1Petersen}
\end{figure}

The metallicity needed to reproduce a frequency ratio of $f_0/f_1$=0.730 with models (Z=0.004, see Figure~\ref{Fig-O1Petersen}), is much larger than the spectroscopic value (Z=0.0002, see also Section~\ref{param}). Also, the average frequency values are quite far ($2f_1-f_0-f_2 = 0.055$) from the resonance condition that could explain the presence of the second overtone by resonant excitation, and which would also have the power to shift the frequency away from the expected value in the Petersen diagram. We note, however, that the frequency values of $f_1$ and $f_2$ are not strictly constant during the observed time span, but undergo irregular fluctuations. We performed a time-dependent Fourier analysis (see also section~\ref{f2stability}) and found the resonance criterion to be fulfilled occasionally. For $f_1$, one combination with $f_0$ and 5 combination frequencies with $f_0$ and $f_B$ were found, leaving the other 6 significant peaks unidentified. One of them, a peak at 2.639 $\rm{d}^{-1}$ (with a frequency ratio of 0.739 with the fundamental mode), would fulfill the resonance criterion, but its amplitude is only 1.9 mmag, compared to 2.8 mmag of $f_1$ . We therefore conclude that the identities of the peaks at 2.6676 $\rm{d}^{-1}$ and 2.639 $\rm{d}^{-1}$ cannot be unambiguously assessed.\\

\begin{table}
\caption{List of the highest peaks in every one of the four regions of excess signal, and their combinations.}
\begin{center}
\begin{tabular}{lcc}
\label{tab-addmodes}
 & frequency & amplitude  \\
 & d$^{-1}$ & mmag \\
\hline
$	f_N	$	&	2.7719	&	3.67		\\
$	f_N+f_0	$	&	4.7211	&	1.58		\\
$	f_N+2f_0	$	&	6.6778	&	0.74		\\
$	f_N+3f_0	$	&	8.6196	&	0.51	\\
$	f_N-f_0	$	&	0.8228	&	0.75		\\
			&		&			\\
$	f_N+f_B	$	&	2.7895	&	1.64		\\
$	f_N+f_0+f_B	$	&	4.7389	&	1.43		\\
$	f_N+2f_0+f_B	$	&	6.6880	&	1.22		\\
$	f_N+3f_0+f_B	$	&	8.6372	&	0.83		\\
$	f_N+4f_0+f_B	$	&	10.5864	&	0.54		\\
$	f_N+5f_0+f_B	$	&	12.5359	&	0.33		\\
$	f_N-f_0-f_B	$	&	0.8044	&	0.69		\\
			&		&		\\
$	f_N+2f_B	$	&	2.8091	&	2.24		\\
$	f_N+2f_0+2f_B	$	&	6.7067	&	0.87		\\
$	f_N+3f_0+2f_B	$	&	8.6559	&	0.57		\\
$	f_N+5f_0+f2f_B	$	&	12.5543	&	0.34		\\
			&		&		\\
$	f_N+2f_0+f_S	$	&	6.6778	&	0.74		\\
$	f_N+2f_0+f_B+f_S	$	&	6.6964	&	0.76	\\
$	f_N+3f_0+f_B+f_S	$	&	8.6456	&	0.81		\\
$	f_N+3f_0+3f_B+f_S	$	&	8.6828	&	0.53		\\
\hline									
$	f_1	$	&	2.6676	&	2.80		\\
$	f_1+f_0	$	&	4.6166	&	1.08		\\
			&		&			\\
$	f_1+f_0+f_B	$	&	4.6362	&	1.16		\\
$	f_1+2f_0+f_B	$	&	6.5851	&	1.02		\\
$	f_1+3f_0+f_B	$	&	8.5343	&	0.77		\\
$	f_1+4f_0+f_B	$	&	10.4838	&	0.45		\\
$	f_1-f_0+f_B	$	&	0.7374	&	1.16		\\
\hline									
$	f_H	$	&	2.9256	&	2.55		\\
$	f_H+f_0	$	&	4.8736	&	0.87		\\
$	f_H+3f_0	$	&	8.7743	&	0.60		\\
$	f_H+4f_0	$	&	10.7234	&	0.43		\\
			&		&			\\
$	f_H-f_B	$	&	2.9070	&	2.29		\\
$	f_H+f_0-f_B	$	&	4.8607	&	0.89		\\
$	f_H+2f_0-f_B	$	&	6.8060	&	0.95		\\
$	f_H+3f_0-f_B	$	&	8.7549	&	0.57		\\
$	f_H+4f_0-f_B	$	&	10.7029	&	0.40		\\
$	f_H-f_0-f_B	$	&	0.9579	&	1.01		\\
$	f_H+f_0+f_B	$	&	4.8924	&	1.00	\\
$	f_H+2f_0+f_B	$	&	6.8413	&	0.71		\\
$	f_H+3f_0+f_B	$	&	8.7905	&	0.58		\\
\hline									
$	f_2	$	&	3.3307	&	1.43	\\
$	f_2+f_0	$	&	5.2781	&	0.85	\\
$	f_2+2f_0	$	&	7.2268	&	0.79		\\
$	f_2+3f_0	$	&	9.1760	&	1.04		\\

\hline
\end{tabular}
\end{center}
\end{table}

The highest peaks in every region, as well as their combinations, are listed in Table~\ref{tab-addmodes}. Altogether, 80 frequencies were found to be significant in the vicinity of the additional peaks: four of these were independent frequencies, 40 were combinations of these independent terms with $f_0$, $f_B$ and/or $f_S$, and 36 were peaks which could not be identified as combinations. Together with the 239 frequencies found in the vicinity of the fundamental mode and its harmonics (see Section~\ref{number}), this results in a total of 319 frequencies included in the analysis. \\

\subsection{Separate analysis of subsets}
\label{subsets}
To study the time-dependent behaviour of the frequency pattern, we used both Period04 and the \textit{time-resolved} feature of SigSpec \citep{reegen07}, which makes it possible to analyse large numbers of data subsets in an automated way. The top panel of Fig.~\ref{f2} illustrates the variation of $f_0$ with time as calculated with SigSpec for overlapping subsets of 15 d. With Period04, we analyzed larger subsets of data, consisting of one or more quarters each. During this process we noticed that in the subset containing Q1 to Q4 (Blazhko period 53.9~d), no deviation of the harmonics as described in section~\ref{dev} is observed. This phenomenon seems to occur only in quarters Q6 and Q7 were the Blazhko periods found in separate analyses of the quarters were 79.8 and 80.4~d, respectively.\\


\subsubsection{Stability of the additional frequencies}
\label{f2stability}
\begin{figure}
\includegraphics[width=90mm, bb= 0 0 455 470]{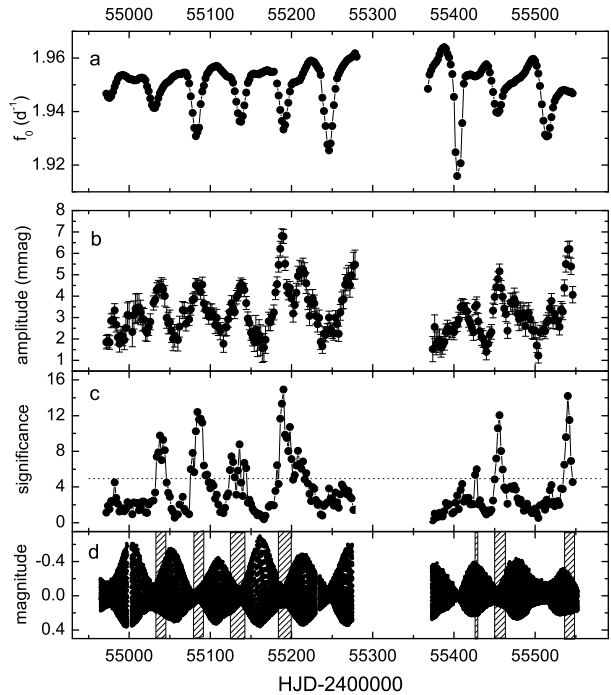}
\caption{Variation of $f_0$ over the complete data set (panel a). Middle panels show the variation of the amplitude (panel b) and the significance (panel c) of the second overtone $f_2$, calculated for bins of 15~d duration with a step width of 2~d. The bottom panel (d) shows the light curve for orientation. The times of significant $f_2$ amplitude are marked with shaded boxes.}
\label{f2}
\end{figure}

The results discussed in Section~\ref{addmodes} hint towards a variability of the amplitude of $f_2$, and we therefore studied the temporal evolution of this peak in detail. Using the time-resolved mode of Sigspec, we performed a Fourier analysis of overlapping bins with a duration of 15~d in steps of 2~d, limiting the frequency range to the region around $f_2$. The resulting amplitudes are plotted in panel b of Fig.~\ref{f2}. While there is a clear variation of the amplitude ranging from 2 mmag to about 7 mmag, no clear periodicity is discernible. We also performed a Fourier analysis on the resulting amplitude curve and found no significant frequency of variability. We note that in Fig.~\ref{f2}, all values of $A(f_2)$ were included, regardless of significance of the frequency. Only bins with an insufficient number of data points and/or bad frequency resolution were discarded. Therefore we provide as a supporting plot the time-dependent significance of the peaks in panel c, where the most commonly used significance criterion of sig$\geq$5 is indicated with a dashed line.

For better orientation, the bottom panel shows the full data set (light curve) of V445~Lyr, and shaded boxes indicate the regions of enhanced $f_2$ amplitude. There seems to be a preference for phases close to Blazhko minima, which are also the phases were the minima of the funadamental frequency $f_0$ occur, but there is no strict rule that can predict a high $f_2$ amplitude. This irregular amplitude variability can explain the numerous peaks around $f_2$ and its combinations which can be seen in the echelle diagram (Fig.~\ref{fig-addechelle}) and which were discussed in section~\ref{addmodes}.\\

We also checked the temporal variations of the frequency values of all the additional frequencies and found, in addition to the variation of $f_0$ which is plotted in panel a of Fig.~\ref{f2}, slight irregular variations of $f_1$ and $f_2$, which lead to an exact parametric resonance during some time intervals in the observed data. This might result in the momentary and transient excitation of $f_2$ which is seen in Fig.~\ref{f2}. We note that from the theoretical point of view, a resonance is necessary to excite the second overtone in this parameter range, because otherwise this overtone would not become unstable.\\

\subsubsection{Variation of the Blazhko modulation parameters}
The modulation parameters are normally used to describe the properties of the Blazhko modulation of a given star. The traditional parameters are $R_k=A_+/A_-$ where $A_+$ and $A_-$ are the amplitudes of the peak on the higher frequency and lower frequency side of the triplet, respectively, and the phase difference $\Delta \varphi_k=\varphi_+-\varphi_-$. The parameter $k$ denotes the harmonic order. Also the asymmetry parameter $Q=(A_+-A_-)/(A_++A_-)$, which was introduced by \citet{alcock03}, is widely used, as well as the power difference $\Delta A^2=A_+^2-A_-^2$. This parameter was recently shown by \citet{szei} to be the physically most meaningful one, as it is directly correlated to the phase difference between the amplitude and the phase modulation components in their model of modulated oscillation.

\begin{figure}
\includegraphics[width=90mm, bb= 0 0 590 360]{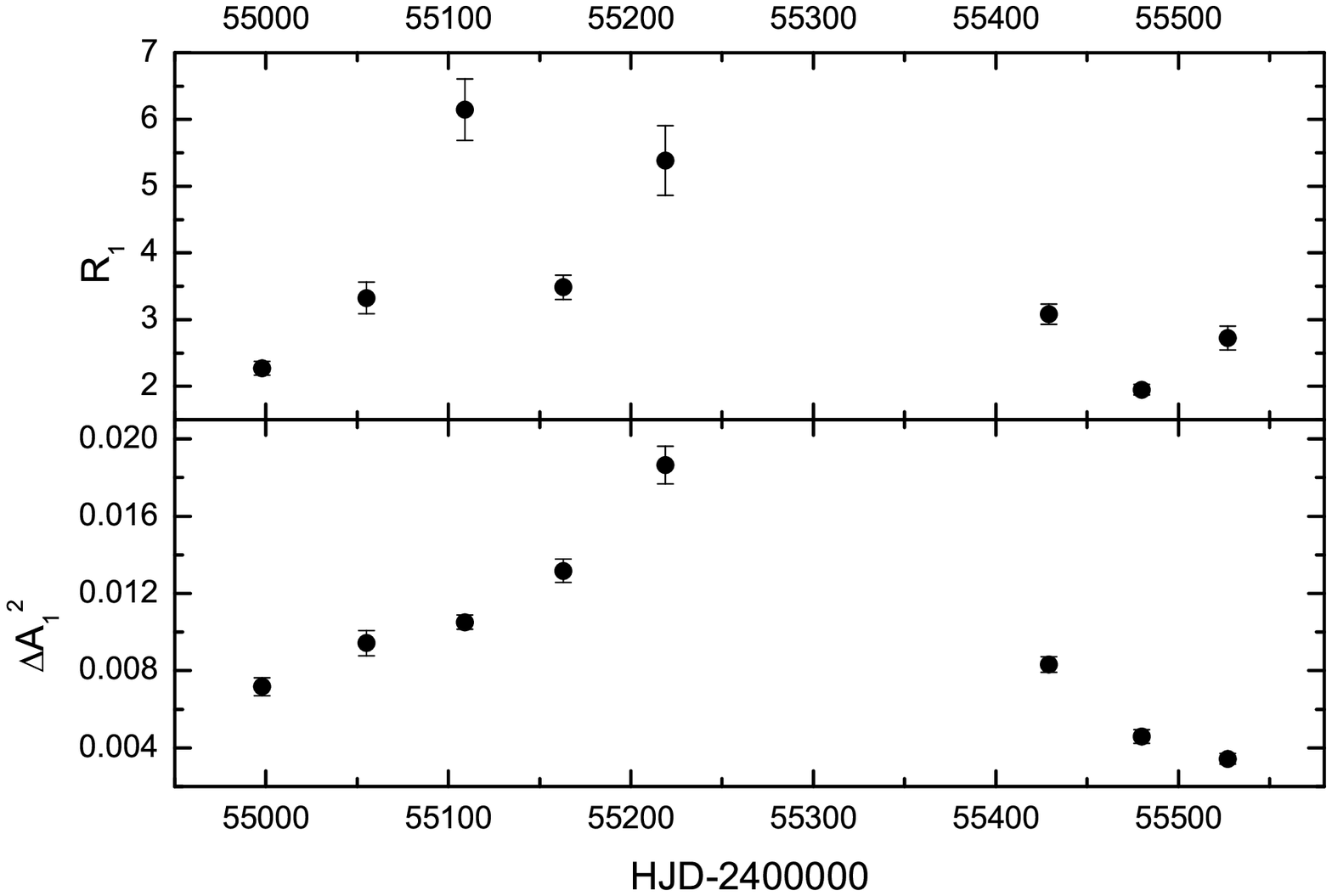}
\caption{Modulation parameters of V445~Lyr versus time. The top panel shows the variation of  $R_1$ ,which is defined as the amplitude ratio between the right and the left modulation side peak, and the bottom panel displays the variation of the power difference $\Delta A_1^2$. The variation of  $\Delta \rm{A_1^2}$ points towards a shift in the phasing between the two types of modulation.}
\label{fig-modparameter}
\end{figure}

\begin{table}
\caption{Overall modulation parameters of V445~Lyr found from the fit to the complete data set for the first 6 harmonic orders.}
\begin{center}
\begin{tabular}{ccccc}
\label{table-modparam}
k	&	$R_k$	&$\Delta \varphi_k$	&$Q_k$	&	$A_k^2$	\\
\hline
1	&	2.112	&	-0.87	&	0.357	&	0.005779	\\
2	&	3.179	&	0.27	&	0.521	&	0.001900	\\
3	&	3.991	&	-0.24	&	0.599	&	0.000445	\\
4	&	3.806	&	-0.08	&	0.584	&	0.000114	\\
5	&	5.382	&	-0.30	&	0.687	&	0.000037	\\
6	&	4.715	&	-0.39	&	0.650	&	0.000011	\\
\hline
\end{tabular}
\end{center}
\end{table}

\begin{figure*}
\includegraphics[width=180mm, bb= 0 0 630 420]{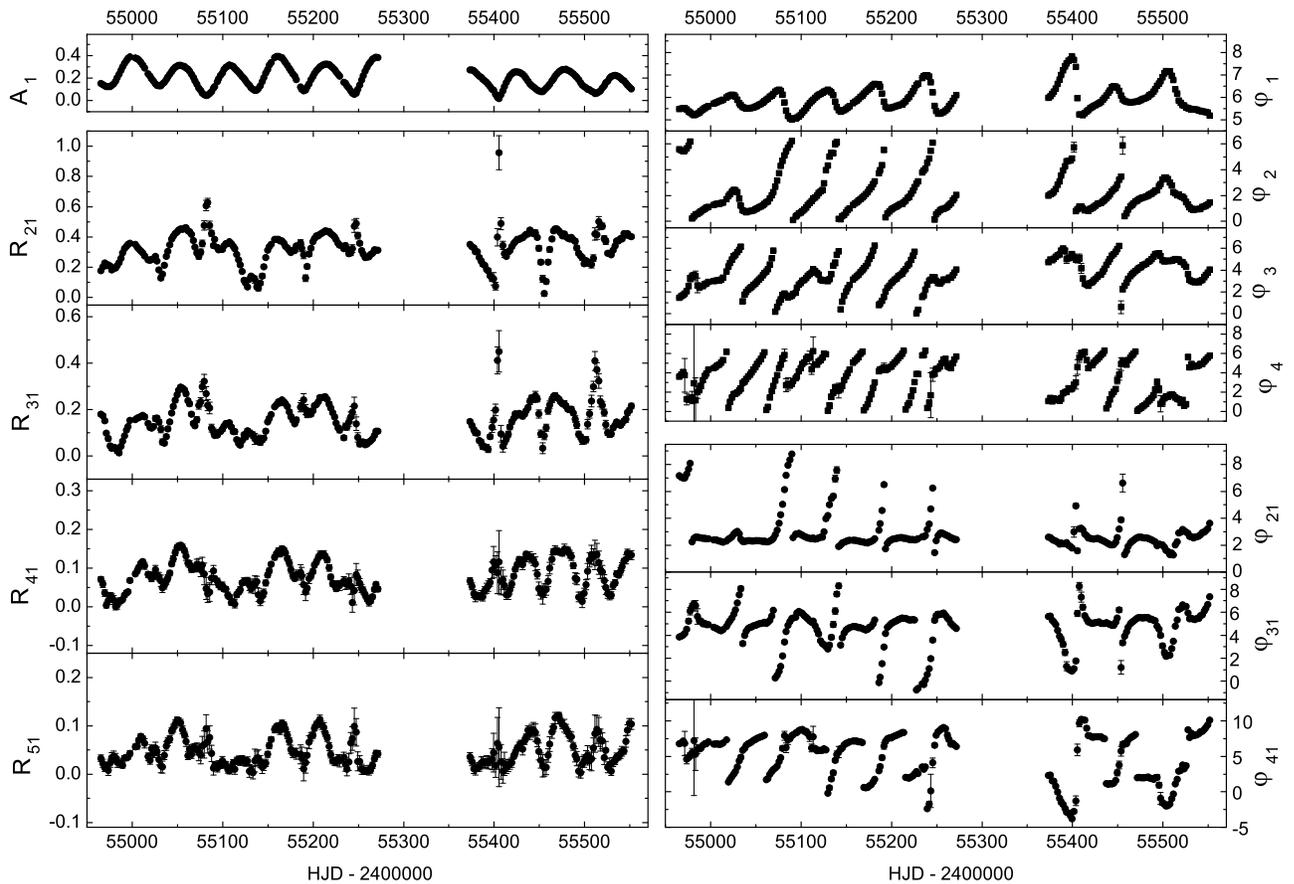}
\caption{Variation of the Fourier parameters during the observed time span, calculated for subsets of 2~d duration each. Left panels: at the top, the amplitude $A_1$ is displayed to indicate where Blazhko maxima and minima are located, while lower panels show the amplitude ratios $R_{21}$,  $R_{31}$, $R_{41}$, and $R_{51}$. Right panels: at the top, the phases $\varphi_{1}$, $\varphi_{2}$, $\varphi_{3}$ and $\varphi_{4}$ are plotted, while the lower panels show the phase differences $\varphi_{21}$, $\varphi_{31}$, and $\varphi_{41}$ in the sine frame. Where error bars are not visible, they are smaller than the symbols.}
\label{fig-fparam}
\end{figure*}

As the main aspect of V445~Lyr is the variability of the Blazhko cycles, we show here not only the average Fourier parameters for every order (which are listed in Table~\ref{table-modparam}) but also their time dependent behaviour in Figure~\ref{fig-modparameter}. To calculate the modulation parameters for every cycle, the data set was divided into bins, starting and ending around Blazhko minimum. The length of the bins was about 60~d, with the exception of the last bin which was only 50~d. Some overlap was allowed to guarantee frequency resolution of the Blazhko multiplet.

It was an important result for the CoRoT star 105288363 that the phasing between the two types of modulation was found to change \citep[see section 5.3 of][]{gug11}. Such a phase change is also indicated by the variation of $\Delta A^2$ of V445~Lyr (bottom panel of Fig.~\ref{fig-modparameter}).


\subsubsection{Fourier parameters}

As the so-called Fourier parameters $R_{k1}$ (amplitude ratio) and $\varphi_{k1}$ (epoch-independent phase difference) are considered useful tools to study and compare the properties of RR Lyrae stars, we calculated their time variability for V445~Lyr. In contrast to the modulation parameters discussed in the previous section, they describe the pulsation rather than the modulation properties. The Fourier parameters are defined as $R_{k1}=A_k/A_1$ and $\varphi_{k1}=\varphi_k-k\varphi_1$. The data were subdivided into 239 bins of 2~d duration, therefore containing about four pulsation periods. On such short timescales, the Blazhko effect is expected to play only a minor role. The effect of period doubling, however, causes a large error in some bins which are more affected by this effect than others. A fit including the fundamental pulsation and 10 harmonics was calculated for each bin, and the results are displayed in Fig.~\ref{fig-fparam} (amplitude ratios are shown in the left panels, phase differences in the right panels). The average values of the parameters (derived from the complete data set) are as given in Table~\ref{tab-fparam}. 

\begin{table}
\caption{Average Fourier parameters of V445~Lyr}
\begin{center}
\begin{tabular}{lclc}
\label{tab-fparam}
parameter & value &parameter & value \\
\hline
$	A_1	$	&	0.184	&	$	\varphi_{1}	$	&	5.84	\\
$	R_{21}	$	&	0.268	&	$	\varphi_{21}	$	&	2.34	\\
$	R_{31}	$	&	0.096	&	$	\varphi_{31}	$	&	4.95	\\
$	R_{41}	$	&	0.033	&	$	\varphi_{41}	$	&	1.08	\\
$	R_{51}	$	&	0.013	&	$	\varphi_{51}	$	&	3.69	\\
\hline
\end{tabular}
\end{center}
\end{table}

Some interesting details are immediately obvious: while the amplitude $A_1$ of 0.18 mag is quite small compared to other RR Lyrae stars, but still in the normal range, the amplitude ratios $R_{k1}$ are significantly smaller than those of non-modulated stars \citep[for comparison, see Fig. 6 in][]{nemec}. An intriguing feature are the sharp upward spikes in the $R_{21}$ variation which occur only during some of the observed Blazhko minima. When looking at the phases one notes that, while $\varphi_1$ has a smooth periodic variation, the phases $\varphi_2$, $\varphi_3$ and $\varphi_4$ show a more or less continuous progression (with exceptions in some cycles), leading to apparent phase jumps in the $\varphi_{k1}$ parameters.

The stability of the results was tested by using also other binnings, and the size of the bins was found to play only a minor role.\\

\subsubsection{Loop diagrams}
\label{loopsect}
\begin{figure}
\includegraphics[width=90mm, bb= 15 5 610 380]{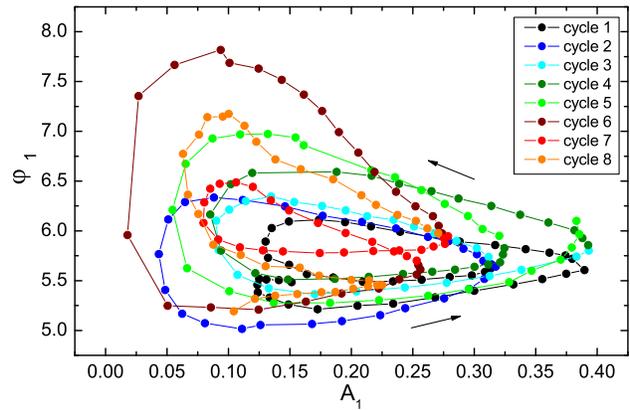}
\caption{$\varphi_1$ versus $A_1$ diagram of V445~Lyr. In this plot it can be seen how the various observed cycles differ from each other. Both the phase modulation component and the amplitude modulation component change over the observed time span. Also, their relative phasing changes from cycle to cycle.}
\label{fig-loop}
\end{figure}

A good indicator for the relative contributions of phase and amplitude modulation, and for the phasing between those two types of modulation, is the $A_1$ versus $\varphi_1$ diagram. The resulting loops for V445~Lyr are plotted in Fig.~\ref{fig-loop}. The direction of motion is indicated with arrows. Cycles are defined for this purpose as from one maximum of $A_1$ to the next, except for the beginning and the end of the data set where the additional points are added to the adjacent cycles. Cycles 1 to 5 correspond to data obtained before the gap in the observations, and cycles 6 to 8 to data after the gap. Even though the contents of Fig.~\ref{fig-loop} are partly redundant with the upper panels of Fig.~\ref{fig-fparam}, this representation allows to better compare the observed Blazhko cycles. All observed Blazhko cycles have a different appearance in this diagram, and the contributions of amplitude and phase modulation change without notable correlation between each other.

\subsection{O-C diagrams and long-term period change}
\label{OC}
\begin{figure}
\includegraphics[width=90mm, bb= 0 0 615 400]{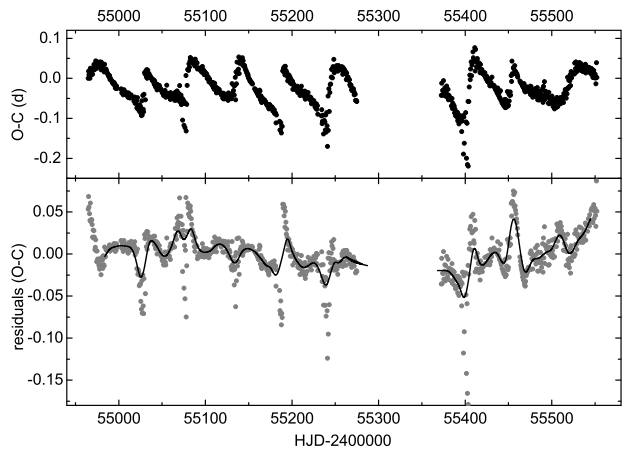}
\caption{O-C diagram of the complete data set of V445~Lyr (top panel), and residuals after subtracting a fit with the Blazhko frequency, its harmonics and the secondary modulation frequency (bottom panel). To guide the eye, a line through the means of the residuals in bins of 10 maxima has been plotted.}
\label{fig-OC}
\end{figure}

In addition to the Fourier analysis which reveals the average Blazhko period, and the time-dependent analysis which shows the fundamental mode as a function of time, we also constructed an O-C diagram, as it can reveal additional details, especially when it comes to long-term period changes. The O-C diagram obtained from all maxima in the \textit{Kepler} V445~Lyr lightcurve is shown in the top panel of Fig.~\ref{fig-OC}. An intriguing feature is the non-sinusoidal variation of the O-C values with a few points at very low O-C values in some of the cycles. A closer inspection of the phase diagrams at the affected times reveals that these drops in O-C are happening at the epochs with the very unusual distorted light curve shape showing double maxima (the double maxima are illustrated in panels b and d of Fig.~\ref{fig-lightcurve}). The determination of the time of the light maximum becomes ambiguous here, depending on the maximum that is chosen. The original maximum seems to move ``to the left" (causing negative O-C values) while a bump on the descending branch gets stronger and takes the role of the maximum for the next Blazhko cycle. This also explains the rather abrupt transition from very low to high O-C values.

We also performed a Fourier analysis on the O-C data to check whether the secondary modulation is also present in the phase variation. We clearly found the Blazhko frequency $f_B$, in this case 0.0184 $d^{-1}$, as well as the harmonics $2f_B$ and $3f_B$ in the O-C curve, which are introduced by the non-sinusoidal variation of the O-C curve. The secondary modulation, with a value of $f_S$ = 0.0064 $d^{-1}$, was also directly detected in the Fourier spectrum, as well as the combination peak $f_B+f_S$. We note that slight differences in the frequency values obtained from the O-C diagram compared to those obtained from the magnitudes do not come unexpectedly, as the phase and amplitude modulations are not strictly correlated.\\
As the quasi-periodic Blazhko modulation is the dominant signal in the O-C curve, it is necessary to subtract a fit including the above mentioned frequencies from the O-C curve to be able to identify long-term changes. The residuals of this fit are shown in the bottom panel of Fig.~\ref{fig-OC} and reveal quite clearly a long-term period change. It remains unclear, however, whether it is a periodic or a continuous linear period change.\\

\subsection{The analytic modulation approach}
\label{analytical}

\begin{figure*} 
\includegraphics[width=17cm, bb= 40 50 750 550]{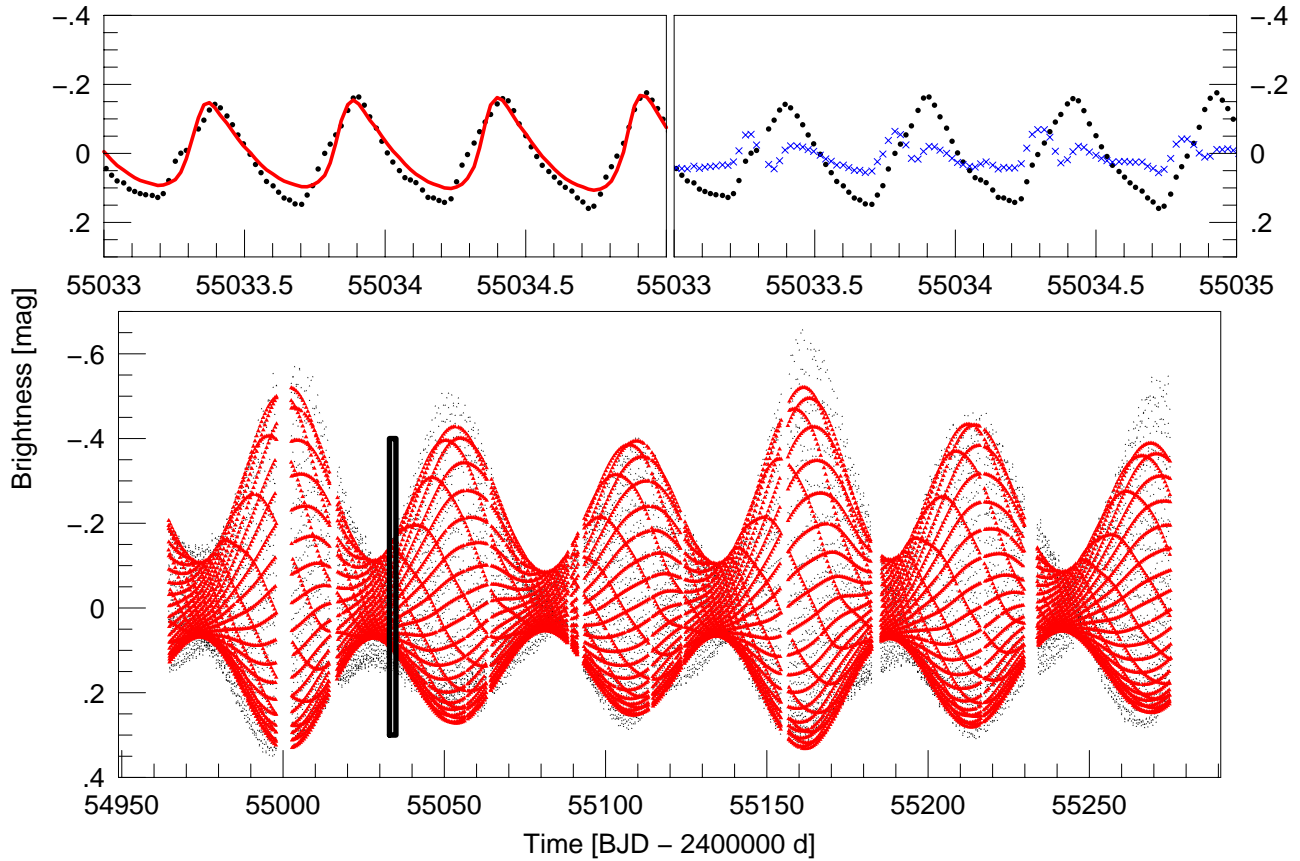}
\caption{
The light curve fit of V445\,Lyr during Q1-Q4 using the modulated signal
approach. The bottom panel shows the global 32-parameter fit (red 
continuous line) with the original data (black dots).
The upper panels show a local fit (left) and residual light curves (blue
crosses) in the interval indicated by the rectangle in the bottom panel.} 
\label{fig_bj}
\end{figure*}
Due to the limitations of the classical Fourier analysis discussed in Section~\ref{fourier}, we applied a new method
of analysis to the \textit{Kepler} data of V445\,Lyr, that was
recently described by \citet[][hereinafter B11]{benko11}. In this
approach, the amplitude and frequency modulations
are treated similarly to the theory of electronic signal
transmitting, reducing dramatically the necessary number
of free parameters. This would be especially useful for stars like V445~Lyr where a classical Fourier analysis yields several hundreds of combination peaks.\\
As a first step of such an analysis we have to select the 
fitting formula, using Table~1  
in B11. The amplitude modulation (AM) with
the frequency of $f_{\mathrm B}$ is evident already from the shape
of the light curve in Fig.~\ref{fig-lightcurve}. Since the envelope curve
is nearly horizontally symmetric for all Blazhko cycles,
the AM of $f_{\mathrm B}$ can be approximated by a simple sinusoidal
function (see formulae 20 and 21 in B11). 

The high asymmetry of the multiplet peaks' amplitudes suggests a 
frequency modulation (FM) as well. Its non-sinusoidal nature is
clear both from the frequency variation function (panel a of Fig.~\ref{f2}) and the
O$-$C diagram (Fig.~\ref{fig-OC}). The combined sinusoidal AM and 
non-sinusoidal FM modulation of $f_{\mathrm B}$ 
can be described by the formula (49) in B11, where $q'=1$.

As mentioned above V445\,Lyr shows a secondary
modulation $f_{\mathrm S}$ as well which is included in the form of an AM because of the changing amplitudes of the
Blazhko cycles. As a first 
approximation we also assumed this modulation as sinusoidal. The linear
combination of $f_{\mathrm B}$ and $f_{\mathrm S}$ shows 
interaction between the two modulations, therefore we have to apply
the formula of modulated modulation (AM cascade -- eq.~27 in B11).

The situation of the FM in $f_{\mathrm S}$ is a bit controversial.
The existence of an FM seems to be well-established, based upon the detection of $f_{\mathrm S}$ in the Fourier analysis of the O-C diagram; however, 
a combined AM with FM in $f_{\mathrm S}$ does not improve the 
significance of our numerical Levenberg-Marquardt fit. 
This may be explained on the basis of the long cycle
length of this modulation and/or its weak FM.

The used best fit model contains a sinusoidal AM and 
a non-sinusoidal FM represented by a three-term Fourier sum 
for the Blazhko modulation and a sinusoidal AM for the second
modulation. The two modulations are assumed to be modulating each other. 
The free parameters are the pulsation frequency $f_0$
and its harmonics' amplitude and phase up to the
9th order ($A_1, A_2, \dots, A_9$, $\varphi_1, \varphi_2, \dots,
\varphi_9$); the modulation frequency $f_{\mathrm B}$, 
the amplitudes and phases of its AM ($a^{\mathrm A}_{\mathrm B 1}$, 
$\varphi^{\mathrm A}_{\mathrm B 1}$) and FM 
($a^{\mathrm F}_{\mathrm{B}i}$, 
$\varphi^{\mathrm F}_\mathrm{Bi}$, $i=1, 2, 3$), the secondary modulation frequency $f_{\mathrm S}$ and
its AM modulation parameters ($a^{\mathrm A}_{\mathrm S 1}$,
$\varphi^{\mathrm A}_{\mathrm S 1}$). They represent 32 parameters
(with the zero point $a_{00}$). The model light curve shows the
global properties of the observed one (see Fig.~\ref{fig_bj}), 
however, the variance of the residual (observed minus fitted) is 
surprisingly high (0.0025 mag).

The large variance may have different reasons. 
Some of them are method-specific, others are
object-specific. An important limit of the method is 
(as mentioned by B11) that it does not describe the 
migration of the humps and bumps caused by the Blazhko effect, a phenomenon which is exceptionally strong in V445~Lyr.
The situation is demonstrated well in the upper panels of Fig.~\ref{fig_bj}.
The other problem is that our method assumes regular signals.
The light curve of V445\,Lyr shows, however, irregular 
behaviour. The loop diagram in Fig.~\ref{fig-loop} illustrates
the cycle-to-cycle variations of the relative strengths of the 
AM and FM components of the modulation.  
Any static (regular) models including the Fourier method 
face similar troubles when using them 
for such a time-dependent (irregular) phenomenon. We conclude that therefore, even though applying the method lead to a success in obtaining a reasonable fit with a comparably small number of parameters, it is nevertheless not optimal for a complicated data set like the one on V445~Lyr.

\section{Spectroscopy, colour photometry and fundamental parameters}
\label{param}

In the framework of ground-based follow up observations, three spectra were obtained with the HIRES spectrograph at the Keck 10~m telescope in August 2011 \citep{nemec12}. Exposure times were 1200~s, and according to the ephemerides given in section~\ref{fulldata} the spectra were obtained shortly after maximum light. Due to its faintness (\textit{Kp}=17.4~mag), V445~Lyr is not an easy target to observe. From a preliminary analysis,  heliocentric radial velocities of -392, -390 and -388 km/s were obtained, as well as a metallicity of [Fe/H]=-2.0 $\pm$ 0.3 dex on the \citet{Carretta} scale, corresponding to Z=0.0003.

\begin{figure}
\includegraphics[width=90mm, bb= 0 0 540 380]{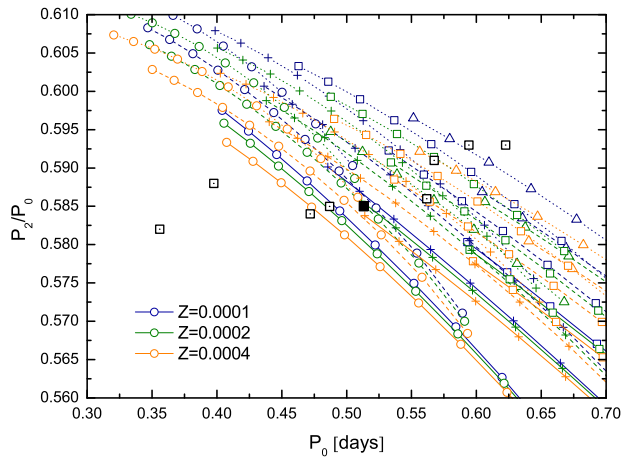}
\caption{Petersen diagram for V445 Lyr (shown as black square) for the second overtone, based on models computed with the Warsaw code \citep{smolec08}. Other RRab Lyr stars with possible second overtones (as discussed in section~\ref{addmodes}) are plotted as open squares.}
\label{fig-petersen}
\end{figure}

Based on the metallicity value derived from spectroscopy and with the frequency value of the second overtone mode, we were able to constrain the mass and luminosity of V445~Lyr with the help of a theoretical Petersen diagram which is shown in Fig.~\ref{fig-petersen}. Linear pulsation models \citep{smolec08} were calculated for a set of masses (0.55, 0.65 and 0.75 $M_\odot$, which are plotted as solid, dashed and dotted lines, respectively) and different luminosities (40, 50, 60, 70 $L_\odot$, plotted as circles, pluses, squares and triangles, respectively). The metallicity values necessary to theoretically fit the observed frequencies agree very well with the measured metallicity, and the luminosity and mass of V445~Lyr found from Fig.~\ref{fig-petersen} are 40-50 $L_\odot$ and 0.55-0.65 $M_\odot$, respectively. 

As the model period ratios depend on the opacity tables and the abundance mixture used in the computations, we tested the stability of our results. The effects of different opacities and mixtures \citep{a09, gn93}, however, were checked and found to play only a minor role. In Fig.~\ref{fig-petersen}, the models based on OPAL opacities and the mixture of \citet{Asplund} are shown. V445~Lyr is plotted as a black square in the diagram, while the other RRab stars for which the presence of the second overtone has been reported, are shown as open squares. From the wide spread which the stars show in Fig.~\ref{fig-petersen}  it is obvious that very different parameters are needed to model the different stars that show the second overtone. Larger masses are neccesary to fit the period ratios of the stars with longer periods, while higher metallicities are needed to to obtain a model for the shorter periods. Models for V350~Lyr and V354~Lyr have  already been shown by \citet{Benko10} in their  Fig. 6, as well as \citet{nemec} in their Fig. A2. As it was already noted in section~\ref{addmodes} that the stars in which the second overtone is excited form a very diverse sample, it is not surprising that they also cover a wide range in mass and metallicity.\\

\subsection{Ground-based color photometry}
To complement the \textit{Kepler} data with colour information on our target, V445~Lyr was observed from the ground in \textit{BVRI} (of which the \textit{RI} bands are in Cousins system) using telescopes at the Lulin Observatory, including the Lulin One-meter Telescope (LOT) and Lulin 0.4m SLT, and the Tenagra II Observatory (TNG, with a 0.8m telescope). The imaging data were reduced with IRAF in a standard manner, including bias and dark subtraction, and flat-fielding. Photometry was obtained from the images using SExtractor \citep{bertin}, and calibrated to the standard magnitudes using standard stars observation from \citet{landolt}. Further details about the telescopes, the CCDs and the reduction of the imaging data can be found in \citet{szabo11} and \citet{ngeow}, who used the same instrumentation for monitoring the Cepheid V1154~Cyg located within the \textit{Kepler} field-of-view \citep{szabo11}. 

The observations were performed between 29 March and 24 July, 2011 (i.e., during the course of two Blazhko cycles) and comprise of about 70 measurements per filter with typical uncertainties of about 0.06~mag. As the data cover the pulsation cycle well, they allow the determination of an average brightness in each colour. The following average magnitudes in the standard system were obtained for V445~Lyr on the basis of the magnitudes of the single measurements: \textit{B}=17.80~mag, \textit{V}=17.38~mag, \textit{R}=17.09~mag and \textit{I}=16.81~mag.\\

\section{Comparison with CoRoT 105288363}
\label{comp-corot}
With the tools developed for the analysis of V445 Lyr, we revisited CoRoT 105288363 to apply the same techniques in a consistent way. Unlike in previous studies, the frequencies were kept as free parameters during the prewhitening and fitting procedure, in spite of the increase of computing time. This has the advantage that the Blazhko period can be determined not only from one measured distance between 2 peaks, but from a large number of independently detected peaks. The standard deviation of that set of measured values also gives a good error estimate. The Blazhko period found by this method is 34.6 $\pm$ 1.1 d and our solution agrees within the error of the previous published value of 35.6 d \citep{gug11}.

\begin{figure}
\includegraphics[width=90mm, bb= 0 0 580 390]{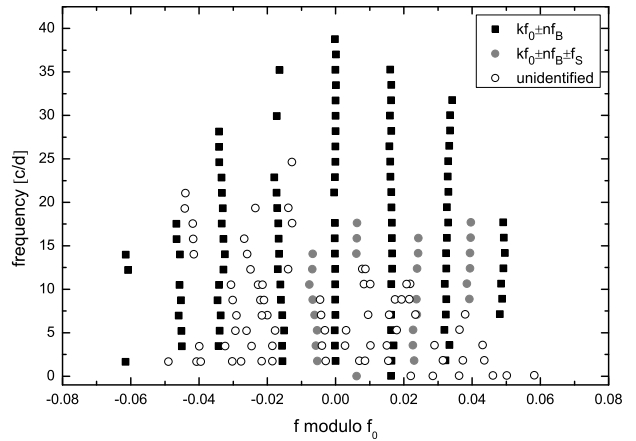}
\caption{Same as Figure~\ref{fig-echelle} but for CoRoT 105288363. Note that the multiplet structure is more symmetrical than in V445 Lyr, in the sense that approximately the same number of multiplet components appear on both sides of the fundamental mode. The peaks belonging to the secondary modulation are in similar positions as in V445 Lyr. Again, several additional peaks occur which cannot be identified as combination peaks of either one of the modulations. This is most likely due to the either irregular or yet unresolved changes of the Blazhko effect. Unlike in V445 Lyr, no combinations including a $2f_S$ term could be found.}
\label{fig-corotechelle}
\end{figure}

\begin{figure}
\includegraphics[width=90mm, bb= 0 0 570 390]{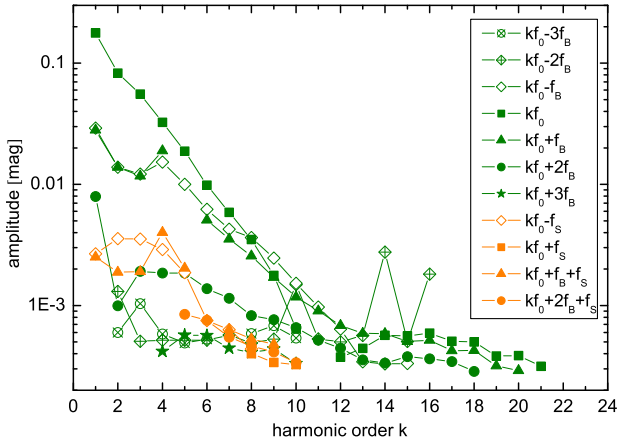}
\caption{Same as Figure~\ref{fig-ampdecr} but for CoRoT 105288363. Here, in contrast to V445 Lyr, the components belonging to the secondary modulation never reach higher amplitudes than the classical multiplet, which might be connected to the fact that the changes in the Blazhko effect of CoRoT 10528836 are not so dramatic as they are in V445 Lyr. Their amplitude decrease is almost as rapid as for the classical components.}
\label{fig-corotampdecr}
\end{figure}

\begin{figure}
\includegraphics[width=90mm, bb= 0 0 540 390]{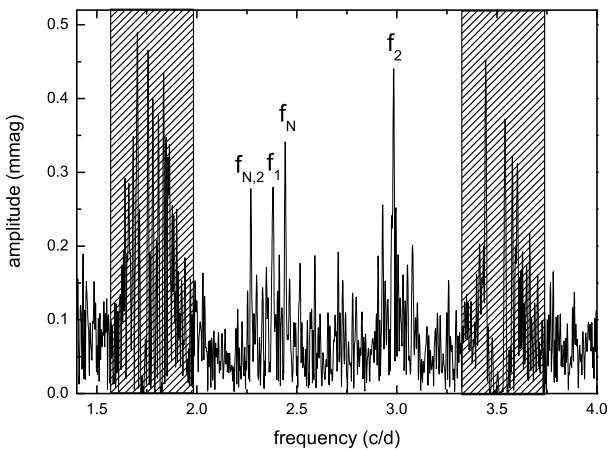}
\caption{Fourier spectrum of CoRoT 105288363 after subtraction of 135 frequencies around the fundamental mode and its harmonics. A zoom into the region between $f_0$ and $2f_0$ is shown, clearly revealing the four additional peaks discussed in the text. The former positions of the multiplets around $f_0$ and $2f_0$ are marked with shaded boxes.}
\label{fig-corotfourier}
\end{figure}

When the echelle diagram diagnostic was applied to CoRoT 105288363, some previously undiscovered features could be unveiled. The first and most important finding is a well-resolved secondary modulation which is very similar to the one in V445 Lyr, in the sense that it has a ratio with the primary modulation of $f_B/f_S$ = 2.5 $\pm$ 0.27, which is close to the value of $f_B/f_S$  = 2.7 $\pm$ 0.12 in V445 Lyr. Also, its combination peaks appear in similar positions: They are clearly found in the higher-frequency side of the harmonics, and also preferentially on the higher-frequency side of the classical Blazhko multiplets. The echelle diagram for  CoRoT 105288363 is shown in Fig.~\ref{fig-corotechelle}. One has to note that, unlike in V445~Lyr, no systematic deviation of the harmonics from their expected positions (see section~\ref{dev}) is found in CoRoT 105288363. Therefore, no tilt of the orders in the echelle diagram can be seen. There are, however, some irregularities in the pattern with some peaks showing deviations from the exact position, though not in a systematic manner. The decrease of amplitudes of the secondary modulation components with increasing harmonic order is shown in Fig.~\ref{fig-corotampdecr}. Those figures should be compared to the corresponding ones for V445 Lyr (Figs.~\ref{fig-echelle} and \ref{fig-ampdecr}, respectively).

The data of CoRoT 105288363 were then also inspected carefully for signal outside the vicinity of the Blazhko multiplets to find evidence for overtones and other possible additional modes. The prewhitening of not only the classical Blazhko multiplet but also the secondary modulation and the additional peaks (which also leads to the disappearance of all aliases) significantly reduced the noise level of the residuals in the frequency region of interest. The second overtone that was reported by \citet{gug11} was also found in the new analysis (see Fig.~\ref{fig-corotfourier}, where it is labeled $f_2$). Due to the reduced noise, also combinations of $f_2$ with $kf_0$ up to $k$=5 could be detected, clearly indicating that the frequency is not due to noise, and not caused by a background star. Moreover, three additional frequencies could be found (see also Fig.~\ref{fig-corotfourier}). A peak at 2.3793 $\rm{d}^{-1}$ with an amplitude of 0.3 mmag is found to be most likely the first overtone due to its ratio of $f_0/f_{1} = 0.741$ with the fundamental mode. Also, a resonance with $f_0$ and $f_2$ is possible in this case, because $2f_1-f_2-f_0$ is only 0.01. Another peak appears at $f_N$=2.4422 $\rm{d}^{-1}$ with an amplitude of 0.35 mmag. It cannot be attributed to any overtone. This and the fact that it appears between the positions of the first and the second overtone makes it quite similar to the observations in V445 Lyr (see section~\ref{addmodes}). Also, its amplitude is slightly higher than that of the suspected first overtone, as is the case for V445 Lyr. The third frequency found in this range is at $f_{N,2}$=2.2699 $\rm{d}^{-1}$ and has an amplitude of 0.3 mmag. It cannot be identified as a radial overtone, and it has no counterpart in V445~Lyr.

Table~\ref{comp} compares the main characteristics of the two stars, including the ratios of the Blazhko modulation, while Table~\ref{comp-addmodes} compares the additional (overtone) modes.

In CoRoT 105288363, there is no sign of period doubling, and no frequencies could be detected at or near the positions of the characteristic half-integer frequencies ($3f_0/2$, $5f_0/2$, etc.).\\

\begin{table}
\caption{Comparison of the main characteristics of the two stars. Fundamental mode frequencies are very different, as well as the Blazhko periods and the periods of the secondary modulation. Interstingly, however, the ratio between the primary and the secondary modulation are very similar.}
\begin{center}
\begin{tabular}{lcc}
\label{comp}
 & CoRoT 105288363 & V445~Lyr\\
\hline
$f_0~\rm{[d^{-1}]}$ & 1.76231 & 1.94903 \\
$P_{0}$ [d] & 0.5674 & 0.5131\\
$P_{B}$ [d] & 34.6 & 53.1\\
$f_{B} \rm{[d^{-1}]}$ & 0.0289& 0.0188\\
$f_{S} \rm{[d^{-1}]}$ &0.0115 & 0.0069\\
$P_{S}$ [d] & 86.5 $\pm$ 9 & 143.3 $\pm$ 5.8\\
\hline
$P_{B}/P_{S}$ &2.5 $\pm$ 0.27 & 2.7 $\pm$ 0.12\\
$P_{B}/P_{0}$ & 60.9 & 103.5\\
\hline
\end{tabular}
\end{center}
\end{table}

\begin{table}
\caption{Comparison of the overtone modes and the additional frequency of the two stars.}
\begin{center}
\begin{tabular}{lcc}
\label{comp-addmodes}
 & CoRoT 105288363 & V445~Lyr\\
\hline
$f_{2} \rm{[d^{-1}]}$ & 2.9856 & 3.3307 \\
$A(f_2) \rm{[mmag]}$ & 0.5 & 1.4 \\
$f_0/f_{2}$ & 0.590 & 0.585\\
\hline
$f_{1} \rm{[d^{-1}]}$ &2.3793 & 2.6676\\
$A(f_1) \rm{[mmag]}$ & 0.3 & 2.7\\
$f_0/f_{1}$ & 0.741 & 0.731 \\
\hline
$f_{N} \rm{[d^{-1}]}$ & 2.442 & 2.7719\\
$A(f_{N}) \rm{[mmag]}$ & 0.4 & 3.6 \\
$f_0/f_{N}$ & 0.722 & 0.703 \\
\hline
\end{tabular}
\end{center}
\end{table}

\section{Summary and Conclusions}
\label{disc}

\subsection{An unusual star with unusual phenomena}
V445~Lyr is an RRab star with such a strong Blazhko modulation that at Blazhko minimum the peak-to-peak amplitude decreases down to 0.07~mag compared to approximately 1~mag during Blazhko maximum (a difference by a factor of 14), leading to the rather low overall amplitude $A_1$ of 0.18~mag. \\
The light curve around Blazhko minimum shows a strong distortion with a secondary maximum, making it impossible to even identify V445~Lyr as an RRab pulsator when observed only during Blazhko minimum.\\
In V445~Lyr the full variety of all the recently discovered new features in RR Lyrae stars -- period doubling, strong irregular changes in the Blazhko effect, a secondary modulation, radial overtone pulsation as well as a non-radial mode -- are combined in one single star. Therefore, it serves as an example of how ultra-precise and uninterrupted space photometry can change our view on seemingly well-known types of stars.
The Fourier phases show an unusual behaviour which has not been detected before in an RR Lyrae star. Also, the distinct spikes in the temporal variation of the Fourier amplitude ratios are a previously unknown feature which is most likely caused by the pronounced double maximum at those phases.

\subsection{Secondary modulation, irregular behaviour, and long-term period change} 
In V445~Lyr, we find the most extreme variations of the Blazhko modulation known so far. This is partly, but not fully, explained by the secondary modulation of 143 d which we found in the \textit{Kepler} data. Irregular/chaotic changes of the Blazhko modulation and/or even longer modulation periods also seem to be present, leading to a dense spectrum of peaks around the harmonics of the fundamental mode, in which with classical methods up to 771 frequencies would be found.\\
The amplitudes of the peaks connected to the secondary modulation were found to decrease less steeply with harmonic order than the components of the classical Blazhko multiplets. This interesting feature yet awaits a physical explanation.\\
A long-term period change is also present, but it could not yet unambiguously be determined whether it is a periodic or a linear change (or neither). Future \textit{Kepler} observations in the upcoming quarters will certainly reveal more about this long-term change.

\subsection{Additional frequencies}
We find at least four additional frequencies not connected to the fundamental mode, its harmonics and the Blazhko peaks. One of these peaks was interpreted as the second radial overtone, one could possibly be the first radial overtone, one was found to be due to period doubling, and the fourth was attributed to a non-radial frequency. \\
The second overtone is not always present during the observations. A strict dependence of the $f_2$ amplitude on the Blazhko phase could not be found. Instead, it seems to vary rather irregularly.\\
Amplitudes and frequencies of all additional peaks change notably during the time span of the data. We consider it possible that fluctuations in the frequency values of $f_1$ lead to transient resonances which temporarily excite the second overtone.\\
The additional peaks form numerous combinations with $f_0$ and the Blazhko multiplets, (including quintuplet peaks) and also with the peaks belonging to the secondary modulation, indicating that they are all intrinsic to the target. Altogether, 80 peaks were found above the significance level at or near the combinations of those 4 frequencies with the other intrinsic frequencies of the target.

\subsection{Spectroscopy, Petersen diagrams, and an alternative method of light curve analysis}
Spectroscopy with the Keck telescope revealed a metallicity of [Fe/H]=-2.0 $\pm$ 0.3, and Petersen diagrams calculated from linear pulsation models point towards a mass of 0.55-0.65 $M_{\odot}$ and a luminosity of 40-50 $L_{\odot}$.\\
We also applied the new analytic modulation technique to the light curve, and found that the best model contains a sinusoidal AM, a non-sinusoidal FM, as well as a sinusoidal AM for the secondary modulation. Due to the migration of a strong bump feature and due to the irregular/stochastic changes, however, the method faces similar troubles as the Fourier analysis.

\subsection{Comparison with another peculiar star}
A revisit of the data on CoRoT 105288363 revealed a secondary modulation period with a similar period ratio (2.5) with the primary modulation period as V445~Lyr (2.7).\\
The new analysis of the CoRoT 105288363 data also points towards the excitation of more additional modes than the previously published second overtone. A non-radial mode as well as the first overtone might also be excited.\\
V445~Lyr also shows a change in the phasing of the two types of modulation (amplitude and phase modulation), similar to what was observed in CoRoT~105288363.

\section*{Acknowledgments}

Funding for this Discovery mission is provided by NASA's Science Mission Directorate. EG acknowledges support from the Austrian Science Fund (FWF), project number P19962-N16. K. Kolenberg is presently a  Marie Curie Fellow (IOF-255267). The research leading to these results has received funding from the European Commission's Seventh Framework Programme (FP7/2007-2013) under grant agreement no. 269194 (IRSES/ASK). RSz and JMB are supported by the 'Lend\"ulet' program of the Hungarian Academy of Sciences and the Hungarian OTKA grants K83790 and MB08C 81013. RSz was supported by the J\'anos Bolyai Research Scholarship of the Hungarian Academy of Sciences. CCN thanks the funding from the National Science Council (of Taiwan) under the contract NSC 98-2112-M-008-013-MY3. We acknowledge the assistance of the queue observers, Chi-Sheng Lin and Hsiang-Yao Hsiao from the Lulin Observatory, and we thank Jhen-kuei Guo and Neelam Panwar for coordinating observations at the Tenagra II Observatory. J.C. and B.S. are grateful to NSF grant AST-0908139 for partial support. Support for MC is provided by the Ministry for the Economy, Development, and Tourism's Programa Inicativa Cient\'{i}fica Milenio through grant P07-021-F, awarded to The Milky Way Millennium Nucleus; by Proyecto Basal PFB-06/2007; by FONDAP Centro de Astrof\'{i}sica 15010003; by Proyecto FONDECYT Regular \#1110326; and by Proyecto Anillo ACT-86.The authors gratefully acknowledge the entire \textit{Kepler} team, whose outstanding efforts have made these results possible.

\label{lastpage}


\begin{thebibliography}{99}
\bibitem[\protect\citeauthoryear{Alcock et al.}{2003}]{alcock03} Alcock C. et al., 2003, ApJ, 598, 597
\bibitem[\protect\citeauthoryear{Arellano Ferro et al.}{2012}]{Arellano} Arellano Ferro A., Bramich D.M., Figuera Jaimes R., Giridhar S., Kuppuswamy K., 2012, MNRAS, 420, 1333
\bibitem[\protect\citeauthoryear{Asplund et al.}{2004}]{Asplund} Asplund M., Grevesse N., Sauval A. J., Allende Prieto C., Kiselman D., 2004, A\&A 417, 751
\bibitem[\protect\citeauthoryear{Asplund et al.}{2009}]{a09} Asplund M., Grevesse N., Sauval A.J., Scott P., 2009, ARA\&A 47, 481
\bibitem[\protect\citeauthoryear{Benk\H{o} et al.}{2010}]{Benko10} Benk\H{o} J.M. et al., 2010, MNRAS 409, 1585
\bibitem[\protect\citeauthoryear{Benk\H{o}, Szab\'o \& Papar\'o}{2011}]{benko11} Benk\H{o} J.M., Szab\'o R., Papar\'o M., 2011, MNRAS, 417, 974
\bibitem[\protect\citeauthoryear{Bertin \& Arnouts}{1996}]{bertin} Bertin E., Arnouts S., 1996, A\&AS, 117, 393
\bibitem[\protect\citeauthoryear{Blazhko}{1907}]{Blazhko07} Blazhko S.N., 1907, AN 175, 325
\bibitem[\protect\citeauthoryear{Buchler \& Koll\'ath}{2011}]{buchler11} Buchler R., Koll\'ath Z., 2011, ApJ 731, 24
\bibitem[\protect\citeauthoryear{Carretta et al.}{2009}]{Carretta} Carretta E., Bragaglia A., Gratton R., D'Orazi, Lucatello S., 2009, A\&A, 508, 695
\bibitem[\protect\citeauthoryear{Catelan}{2009}]{catelan09} Catelan M., 2009, Ap\&SS, 320, 261
\bibitem[\protect\citeauthoryear{Celik et al.}{2011}]{celik11} Celik L. et al., 2011, proceedings in prep
\bibitem[\protect\citeauthoryear{Chadid et al.}{2010}]{cha} Chadid M. et al.., 2010, A\&A 510, 39
\bibitem[\protect\citeauthoryear{Christiansen et al.}{2011}]{christ} Christiansen J. L. et al., 2011, \textit{Kepler} Data Characteristics Handbook (KSCI-19040-002)
\bibitem[\protect\citeauthoryear{Dziembowski \& Cassisi}{1999}]{DC99}  Dziembowski W., Cassisi S., 1999, AcA, 49, 371
\bibitem[\protect\citeauthoryear{Grevesse \& Noels}{1993}]{gn93} Grevesse N., Noels A., 1993, in Origin and Evolution of the Elements, Cambridge Univ. Press, Cambridge, p. 15
\bibitem[\protect\citeauthoryear{Guggenberger et al.}{2011}]{gug11} Guggenberger E., Kolenberg K., Chapellier E., Poretti E., Szab\'o R., Benk\H{o} J. M., Papar\'o M., 2011, MNRAS 415, 1577
\bibitem[\protect\citeauthoryear{Haas et al}{2010}]{haas10} Haas M., 2010, ApJ 713, L115
\bibitem[\protect\citeauthoryear{Jenkins et al.}{2010}]{jen10} Jenkins J.M. et al., 2010, ApJ 713, L87
\bibitem[\protect\citeauthoryear{Jurcsik et al.}{2005}]{jur05} Jurcsik J. et al., 2005, A\&A 430, 1049
\bibitem[\protect\citeauthoryear{Jurcsik et al.}{2008}]{jur08} Jurcsik J. et al., 2008, MNRAS 391, 164
\bibitem[\protect\citeauthoryear{Jurcsik et al.}{2009}]{jur09c} Jurcsik J. et al., 2009, MNRAS 400, 1006
\bibitem[\protect\citeauthoryear{Jurcsik et al.}{2012}]{jur12} Jurcsik J. et al., 2012, MNRAS, submitted
\bibitem[\protect\citeauthoryear{Kolenberg et al.}{2009}]{kol09} Kolenberg K. et al., 2009, MNRAS 396, 263
\bibitem[\protect\citeauthoryear{Kolenberg et al.}{2010}]{kol10a} Kolenberg K. et al., 2010, ApJ 713, 198
\bibitem[\protect\citeauthoryear{Kolenberg et al.}{2011}]{kol10b} Kolenberg K. et al., 2011, MNRAS, 411, 878
\bibitem[\protect\citeauthoryear{Koll\'ath et al.}{2011}]{kollath11} Koll\'ath Z., Moln\'ar L., Szab\'o R., 2011, MNRAS, 414, 1111
\bibitem[\protect\citeauthoryear{Koch et al.}{2010}]{koch10} Koch D.G. et al., 2010, ApJ 713, L79
\bibitem[\protect\citeauthoryear{Kukarkin et al.}{1973}]{kukarkin73} Kukarkin B.V., Kholopov P.N., Kukarkina N.F., Perova N.B., 1973, IBVS 834
\bibitem[\protect\citeauthoryear{Landolt}{2009}]{landolt} Landolt A.~U, 2009, AJ, 137, 4186
\bibitem[\protect\citeauthoryear{Lenz \& Breger}{2005}]{Lenz} Lenz P., Breger M., 2005, CoAst 146, 53
\bibitem[\protect\citeauthoryear{Nemec et al.}{2011}]{nemec} Nemec J. M. et al. 2011, MNRAS 417, 1022
\bibitem[\protect\citeauthoryear{Nemec, Cohen \& Sesar}{2012}]{nemec12} Nemec J. M., Cohen J. G., Sesar B., 2012, in preparation
\bibitem[\protect\citeauthoryear{Ngeow}{2011}]{ngeow} Ngeow C.-C., 2011, Submitted for the 9th Pacific Rim Conference on Stellar Astrophysics (PRCSA2011), arXiv:1111.2094
\bibitem[\protect\citeauthoryear{Poretti et al.}{2010}]{poretti} Poretti E. et al. 2010, A\&A, 520, A108
\bibitem[\protect\citeauthoryear{Popielski, Dziembowski \& Cassisi}{2000}]{popielski} Popielski B.L., Dziembowski W.A., Cassisi S., 2000, AcA, 50, 491
\bibitem[\protect\citeauthoryear{Reegen}{2007}]{reegen07} Reegen P. 2007, A\&A 467, 1353
\bibitem[\protect\citeauthoryear{Romano}{1972}]{romano72} Romano G. 1972, IBVS 645, 1
\bibitem[\protect\citeauthoryear{Shapley}{1916}]{shapley16} Shapley H., 1916, ApJ 43, 217
\bibitem[\protect\citeauthoryear{Smolec et al.}{2011}]{smo11} Smolec R., Moskalik P., Kolenberg K., Bryson S., Cote M. T., Morris R. L., 2011, MNRAS 414, 2950
\bibitem[\protect\citeauthoryear{Smolec  \& Moskalik}{2008}]{smolec08} Smolec R., Moskalik P., 2008, AcA 58, 193
\bibitem[\protect\citeauthoryear{S\'odor et al.}{2011}]{sodor11} S\'odor A. et al., 2011, MNRAS, 411, 1585
\bibitem[\protect\citeauthoryear{S\'odor et al.}{2012}]{Sodor12} S\'odor A. et al., 2012, submitted, arXiv:1201.5474v1
\bibitem[\protect\citeauthoryear{Soszynski et al.}{2009}]{so09} Soszynski I. et al., 2009, AcA 59, 1
\bibitem[\protect\citeauthoryear{Soszynski et al.}{2011}]{Soszynski11} Soszynski I. et al., 2011, AcA 61, 1
\bibitem[\protect\citeauthoryear{Stothers}{2006}]{sto06} Stothers R., 2006, ApJ 652, 643
\bibitem[\protect\citeauthoryear{Szab\'o, Koll\'ath \& Buchler}{2004}]{szabo04} Szab\'o R., Koll\'ath Z., Buchler J. R., 2004, A\&A, 425, 627
\bibitem[\protect\citeauthoryear{Szab\'o et al.}{2010}]{szabo10} Szab\'o R. et al., 2010, MNRAS 409, 1244
\bibitem[\protect\citeauthoryear{Szab\'o et al.}{2011}]{szabo11} Szab{\'o}, R. et al., 2011, MNRAS, 413, 2709
\bibitem[\protect\citeauthoryear{Szeidl \& Jurcsik}{2009}]{szei} Szeidl B., Jurcsik J., 2009, CoAst 160, 17
\bibitem[\protect\citeauthoryear{Van Cleve \& Caldwell}{2009}]{vancleve} Van Cleve J., Caldwell D.A., 2009, \textit{Kepler} Instrument Handbook, KSCI 19033
\bibitem[\protect\citeauthoryear{Van Hoolst, Dziembowski \& Kawaler}{1998}]{vanhoolst} Van Hoolst T, Dziembowski W.A., Kawaler S.D., 1998, MNRAS 297, 536
\end{thebibliography}
\end{document}